\documentclass[12pt,manuscript]{aastex}

\usepackage{natbib}
\usepackage{lscape}
\usepackage{geometry}
\begin{document}

\title{Is the Young Star RZ Piscium Consuming Its Own (Planetary) Offspring?}

\author{K. M. Punzi\altaffilmark{1}, J. H.\ Kastner\altaffilmark{1}, C. Melis\altaffilmark{2}, B. Zuckerman\altaffilmark{3}, C. Pilachowski\altaffilmark{4}, L. Gingerich\altaffilmark{1,5}, T. Knapp\altaffilmark{1,6}}

\altaffiltext{1}{Center for Imaging Science, School of Physics \& Astronomy, and 
Laboratory for Multiwavelength Astrophysics, Rochester Institute of
Technology, 74 Lomb Memorial Drive, Rochester NY 14623, USA}
\altaffiltext{2}{University of California, San Diego, 9500 Gilman Drive, La Jolla, CA 92093, USA}
\altaffiltext{3}{University of California, Los Angeles, CA 90095, USA}
\altaffiltext{4}{Indiana University, 107 S. Indiana Avenue, Bloomington, IN 47405-7000, USA}
\altaffiltext{5}{Haverford College, 370 Lancaster Avenue, Haverford, PA 19041, USA}
\altaffiltext{6}{Ithaca College, 953 Danby Road, Ithaca, NY 14850, USA}

\begin{abstract}
The erratically variable star RZ Piscium (RZ\,Psc) displays extreme optical dropout events and strikingly large excess infrared emission.
To ascertain the evolutionary status of this intriguing star, we obtained observations of RZ\,Psc with the European Space Agency's \textit{X-ray Multi-Mirror Mission} (\textit{XMM-Newton}), as well as high-resolution optical spectroscopy with the Hamilton Echelle on the Lick Shane 3\,m telescope and with HIRES on the Keck\,I 10\,m telescope.
The optical spectroscopy data demonstrate that RZ\,Psc is a pre-main sequence star with an effective temperature of 5600 $\pm$ 75 K and log g of 4.35 $\pm$ 0.10.
The ratio of X-ray to bolometric luminosity, $\log{L_X/L_{\rm bol}}$, lies in the range $-3.7$ to $-3.2$, consistent with ratios typical of young, solar-mass stars, thereby providing strong support for the young-star status of RZ\,Psc.
The Li absorption line strength of RZ\,Psc suggests an age in the range 30--50 Myr, which in turn implies RZ\,Psc lies at a distance of $\sim$170 pc.
Adopting this estimated distance, we find the Galactic space velocity of RZ\,Psc to be similar to the space velocities of stars in young moving groups near the Sun.
Optical spectral features indicative of activity and/or circumstellar material are present in our spectra over multiple epochs, which provide evidence for the presence of a significant mass of circumstellar gas associated with RZ\,Psc.
We suggest that the destruction of one or more massive orbiting bodies has recently occurred within 1 au of the star, and we are viewing the aftermath of such an event along the plane of the orbiting debris.
\end{abstract}

\section{Introduction}

What is the nature of the enigmatic field star RZ\,Piscium (RZ\,Psc)?
This curious case has been debated for decades \citep{2014A&A...563A.139P}.
There exists significant evidence that RZ\,Psc may be a young main sequence star with a dusty disk that is in the late stages of planet formation (\citealt{2013A&A...553L...1D, 2015AstL...41..407G}, and references therein).
The star's sharp dropouts in optical magnitude and its large IR excess have prompted the suggestion that a collisionally active asteroid belt is present in orbit around the star \citep{2013A&A...553L...1D,2017RSOS....460652K}.
Such a scenario is similar to that proposed to explain UX Orionis objects, a class of young stars whose photometric variability is caused by variable circumstellar extinction \citep{1994AJ....108.1906H}.

Indeed, most age estimates for RZ\,Psc are based on the assumption that RZ\,Psc is a UX Ori-type object.
The age of RZ\,Psc was estimated from the equivalent width (EW) of the lithium 6708 \AA\ line as $\sim$10-70 Myr, intermediate between the median ages of stars in the Pleiades and Orion clusters \citep[][]{2010A&A...524A...8G}.
\citet{2013AstL...39..776P} estimated an age of 25 $\pm$ 5 Myr and concluded that the disk orbiting RZ\,Psc is in transition from primordial disk to debris disk.

However, it is worth considering whether RZ Psc might be an evolved star, given that aperiodic optical dropouts are also observed in such cases \citep[e.g., WD 1145+017;][]{2015Natur.526..546V}.
Indeed, RZ\,Psc's high Galactic latitude ($b \cong 35^{\circ}$) and lack of association with any known young stellar group might suggest evolved star status.
If RZ\,Psc is an evolved giant, its disk may be a consequence of the destruction of a low-mass stellar companion or massive planet when the star expanded to become a red giant.
\citet{2009PhDT........34M} chose to call such dusty, first ascent giant stars ``Phoenix Giants'' because of their similarities to T Tauri stars \citep[specifically, high Li abundance and strong infrared excess;][]{2014A&A...563A.139P}.

Furthermore, the strong IR excess associated with RZ\,Psc ($L_{\rm IR}$/$L_{\rm bol}$ $\sim 8\%$, \citealt{2013A&A...553L...1D}) would be very peculiar if the star is as old as $\sim$25 Myr.
Typically, only cloud-embedded pre-main sequence stars have IR excesses this large \citep[Figure 1 in][]{2009AIPC.1158....3M}.
Many hundreds of young and main sequence solar-type stars have been studied over the past few decades (since the launch of the Infrared Astronomical Satellite), yet few have been found to have $L_{\rm IR}/L_{\rm bol}$ as large as or larger than that of RZ\,Psc \citep[examples are TYC 8241 2651 1 and V488 Per;][]{2012Natur.487...74M,2012ApJ...752...58Z}.
However, while RZ\,Psc's level of IR excess is very unusual for a dusty debris disk, it is quite common for a Phoenix Giant \citep{2008ApJ...683.1085Z,2009ApJ...696.1964M,2009PhDT........34M}.

Attempts to distinguish between these evolutionary possibilities -- i.e., a young star with a protoplanetary disk versus an evolved star with disk derived from the destruction of a companion -- can be quite problematic.
X-ray observations may provide key evidence about stellar properties that can break this degeneracy, as has been demonstrated in the case of BP\,Psc \citep{2010ApJ...719L..65K}.
BP\,Psc was initially classified as a classical T Tauri star based on its strong H$\alpha$ and forbidden line emission \citep{1986ApJ...300..779S}, although it is located at high galactic latitude, far from any known star-forming regions.
\citet{2008ApJ...683.1085Z} discovered an orbiting, dusty circumstellar disk and a parsec-scale system of highly collimated outflows, which is consistent with the classification of BP\,Psc as a pre-main sequence star.
However, the location of BP\,Psc in the sky, its weak lithium 6708 \AA\ line, its gravity-sensitive photospheric absorption, the \textit{Spitzer} Infrared Spectrograph spectrum of highly crystalline sub-micron-sized dust grains in its circumstellar disk \citep{2010ApJ...724..470M}, and the millimeter-wave molecular spectrum of its circumstellar disk are atypical for pre-main sequence (pre-MS) star-disk systems \citep{2008A&A...486..239K}.
\textit{Chandra} X-ray observations of BP\,Psc revealed that the star is a weak X-ray point source and its X-ray luminosity ratio ($\log{(L_X/L_{\rm bol})}$ $\sim$ -5.8 to -4.2) lies in the range that is observed for rapidly rotating (FK Com-type) G stars \citep{2010ApJ...719L..65K}.
Hence, the \textit{Chandra} results favor a scenario wherein the disk/jet system of BP\,Psc is the result of a very recent engulfment of a companion as the star ascended the giant branch.

To investigate whether the X-ray properties of RZ\,Psc might similarly discriminate between these potential competing models for its evolutionary state --- i.e., young versus evolved star --- we have obtained observations of the star with \textit{XMM-Newton}.
To further aid in determining the nature of the RZ\,Psc system, we obtained optical spectroscopic observations with the Hamilton Echelle on the Lick Shane 3\,m telescope and with HIRES on the Keck\,I 10\,m telescope.
The observations and data reduction are described in Section~\ref{sec:Observations}, the results and analysis are described in Section~\ref{sec:resultsanalysis}, a discussion of the results is presented in Section~\ref{sec:discussion}, and a summary of the results and an outlook for future work is presented in Section~\ref{sec:summary}.

\section{Observations}
\label{sec:Observations}

\subsection{\textit{XMM-Newton} European Photon Imaging Cameras (EPIC)}

RZ\,Psc was observed by \textit{XMM-Newton} \citep{2001A&A...365L...1J} for $\sim$38.8\,ks on 2015 January 2--3 during revolution 2759.
The observations were conducted with the pn, MOS1, and MOS2 CCD arrays of the European Photon Imaging Cameras (EPIC) \citep{2001A&A...365L..18S,2001A&A...365L..27T} operating in full frame mode.

We also obtained simultaneous optical and UV observations with the Optical Monitor available on \textit{XMM-Newton}.
A summary of the \textit{XMM-Newton} observations and exposure times is listed in Tables~\ref{tbl:XRayObs} and \ref{tbl:OMObs}.
Processing of the raw event data was performed, using standard methods, via the \textit{XMM-Newton} Science Analysis System (SAS version 14.0). 

\begin{figure}[h!]
  \centering
  \includegraphics[width=0.9\hsize]{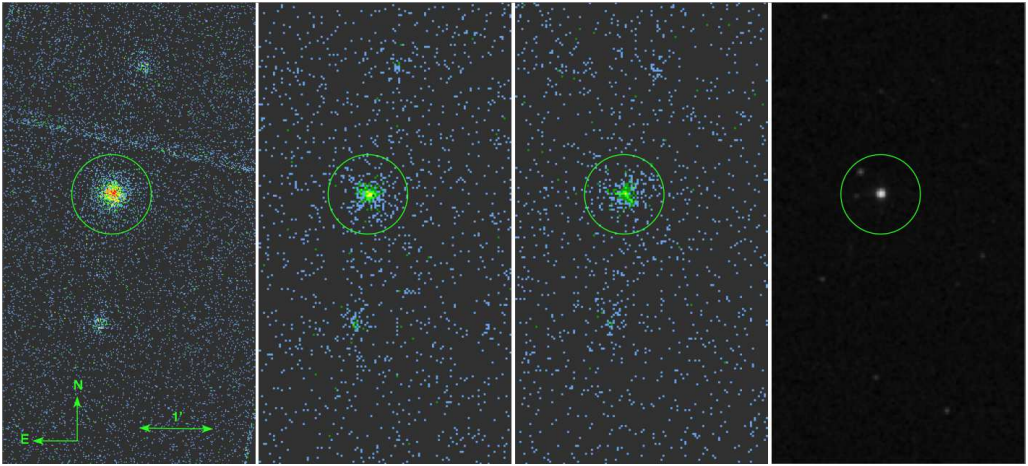}
  \caption{\textit{XMM-Newton} pn (left), MOS1 (middle left), and MOS2 (middle right) images of RZ\,Psc, with orientation and scale indicated, alongside 2MASS J band image (right) at the same orientation and scale.}
  \label{fig:XRayIR}
\end{figure}

In Figure~\ref{fig:XRayIR}, we present the merged EPIC pn, MOS1, and MOS2 X-ray images centered near the position of the target star.
RZ\,Psc is the brightest of the few dozen X-ray sources detected in the observation.
The associated X-ray source lies at  $01^{\rm h}09^{\rm m}42^{\rm s}.05$, $27^{\circ}57^{\prime}01~95^{\prime\prime}$, coincident with the near-IR position of RZ\,Psc (Figure~\ref{fig:XRayIR}, right).

\subsection{High-Resolution Optical Spectroscopy}

\subsubsection{Lick Shane 3\,m Hamilton Echelle}

RZ\,Psc was observed with the Shane 3\,m telescope and Hamilton echelle spectrograph (\citealt{1987PASP...99.1214V}) at Lick Observatory.
The observation dates and parameters are listed in Table~\ref{taboptspec}.
Reduction of the Hamilton echelle data was performed using IRAF tasks following methods outlined in detail in Lick Technical Report No. 74\footnote{See http://astronomy.nmsu.edu/cwc/Software/irafman/manual.html.}.
Briefly, the spectral images were bias-subtracted, flat-fielded, extracted, and finally wavelength-calibrated with TiAr arclamp spectra (see \citealt{2013AJ....146...97P}).

\subsubsection{Keck~I 10\,m HIRES}

RZ\,Psc was observed with the Keck~I 10\,m telescope at Maunakea Observatory, where echelle spectroscopy was performed with the HIRES instrument \citep{1994SPIE.2198..362V}.
Observation dates and parameters are listed in Table~\ref{taboptspec}.
All of the HIRES data were reduced with the {\sf MAKEE} software package.
The spectral images were bias-subtracted and flat-fielded, then the spectra were extracted and finally wavelength-calibrated with ThAr arclamp spectra.

\subsection{Keck~II 10\,m Echellette Spectrograph and Imager (ESI)}

A candidate M-type companion to RZ\,Psc (see\S~\ref{sec:companion}) was observed at Maunakea Observatory with the Keck~II 10\,m telescope.
Echelle spectroscopy was performed with the ESI (\citealt{2002PASP..114..851S}), and observation parameters are given in Table~\ref{taboptspec}.

Data were reduced with standard IRAF tasks and procedures similar to the reduction of Hamilton data as described above.
Specifically, spectral images were bias-subtracted, flat-fielded with the use of dome flat exposures, extracted, and wavelength-calibrated with the use of CuArXeHgNe arc lamps.

\subsection{WIYN-0.9\,m Half Degree Imager (HDI)}

RZ\,Psc and its field were observed at Kitt Peak National Observatory with the WIYN 0.9\,m telescope\footnote{The WIYN Observatory is a joint facility of the University of Wisconsin-Madison, Indiana University, the National Optical Astronomy Observatory and the University of Missouri.} in 2016 January.
The HDI was used to take 10 s exposures of the RZ\,Psc field in BVRI filters.

Data were reduced with standard AstroImageJ tasks and procedures \citep{2017AJ....153...77C}.
Multi-aperture differential photometry was then performed to determine the magnitude of stars in the field in all bands.

\section{Results and Analysis}
\label{sec:resultsanalysis}

In Table~\ref{tab:RZPscCharPrev}, we present the key characteristics of RZ\,Psc that we have gleaned from the literature, and in Table~\ref{tab:RZPscCharNew}, we present the key characteristics determined via the analysis described next.

\subsection{Model Atmosphere Analysis} 
\label{sec:Atmosphere}

An LTE, 1D model atmosphere analysis of the optical spectra of RZ\,Psc obtained with the Keck and Shane telescopes was performed using the spectrum synthesis program Moog \citep[][2014 version]{1973ApJ...184..839S} to determine the atmospheric parameters and composition of the star.
Model atmospheres have been interpolated within the MARCS4 grid \citep{2008A&A...486..951G}.
The line list of \citet{2014MNRAS.441.3396Y} was adopted for atomic data.

Equivalent widths (EWs) were first measured for individual spectra using the \textit{splot} task in IRAF\footnote{IRAF is distributed by the National Optical Astronomy Observatory, which is operated by the Association of Universities for Research in Astronomy (AURA) under a cooperative agreement with the National Science Foundation.}.
Measurements from different epochs were compared to identify any systematic differences in line strength at different epochs.
Since no variations in EWs were found, seven spectra covering a range of dates from 2013 through 2016 were co-added to produce a spectrum of higher S/N ratio; the combined spectrum has a S/N ratio of 200.
The combined spectrum with lower noise also reduces the uncertainty in the continuum level, allowing more accurate determination of EWs.
The final EWs used for the analysis were measured from the combined spectra.

The first step was to carry out the determination of atmospheric parameters -- effective temperature, surface gravity, microturbulence, and metallicity -- independently of previous determinations.
We selected 101 lines of Fe I to cover a range of both line strength and excitation potential, and model atmosphere parameters were varied to minimize the dependence of derived abundance on EW and excitation potential.
The procedure yielded an effective temperature of $5600~\pm~75$ K and a microturbulence of v$_{T}$~=~($2.0~\pm~0.2$)~km s$^{-1}$.
The surface gravity of the star was determined to be $\log{g}$~=~$4.35~\pm~0.10$ by minimizing the difference in the Fe abundance derived from Fe I and Fe II lines.
The resulting metallicity is [Fe/H]~=~-0.11~$\pm$~0.03 (standard error of the mean or S.E.M.), not including systematic errors in the atmospheric parameters.

Abundances of additional species were determined once the model atmosphere parameters were established.
Again, the \citet{2014MNRAS.441.3396Y} line list provided atomic data.
The derived abundances are included in Table~\ref{table:abundances}, along with estimated uncertainties, including both random and systematic errors, where the solar abundances are taken from \citet{2009ARA&A..47..481A}.
The uncertainties in the atmospheric parameters are determined from the parameter change needed to establish a clear dependence of abundance versus excitation potential or abundance versus line strength, or a difference in abundance as derived from Fe I and Fe II lines.

Atmospheric parameters for RZ\,Psc have previously been reported by \citet{2000ARep...44..611K} and by \citet{2014A&A...563A.139P}.
\citeauthor{2000ARep...44..611K} reported a weighted average temperature of 5250~K based on their analysis of three spectra with S/N ratios in the range 44-68 and a spectral resolving power of R~$\approx$~25,000.
They employed three alternate methods, including an unreddened color-temperature relation, specific line ratios, and, as we used here, a minimization of the dependence of abundance on excitation potential for Fe I lines, the ``Boltzman equilibrium."
For this last method, they derived an effective temperature of 5450~K~$\pm$~150~K, which is statistically indistinguishable from our best fit temperature of 5600~$\pm$~75~K.
Although their result for microturbulance (v$_{T}$~=~2.0~$\pm$~0.5~km~s$^{-1}$) agrees with ours, their determination of $\log{g}$~=~3.41~$\pm$~0.02 is discrepant.
   
\citet{2014A&A...563A.139P} analyzed a higher-resolution spectrum (R~$\approx$~46,000) with a S/N ratio of about 75.
Their temperature of 5350~K~$\pm$150~K was obtained from synthetic spectrum fits to a few features, most notably H$\alpha$ and H$\beta$, which may be compromised by circumstellar emission.
With the temperature fixed at 5350~K, they used SME analysis software \citep{1996A&AS..118..595V} to determine $\log{g}$~=~4.2~$\pm$~0.3 and v$_{T}$~=~1 km/s.

To attempt to resolve the discrepancies among these model fits, we determined the elemental abundances of RZ\,Psc using each of the three \citet{2000ARep...44..611K} models with our measured EWs.
The \citeauthor{2000ARep...44..611K} atmospheric parameters provide a consistent abundance for the Fe I and Fe II lines but significant deficiencies in the abundances of the ionized species Sc II and Ti II.
We note that the abundances listed in their Table 3 include large deficiencies of Ti, V, Cr, and Ni, as well as Si and Al.

The \citeauthor{2014A&A...563A.139P} model atmosphere parameters with our EWs suggest that a value of 2~km s$^{-1}$ for the microturbulence provides a better fit.
With that correction, adopting the \citeauthor{2014A&A...563A.139P} model leads to a significant discrepancy ($\Delta$[Fe/H]~=~0.2 dex) between the Fe I and Fe II lines, as well as a low abundance of Fe compared to other metals.

Our adopted model parameters provide both a good match for the Fe I and Fe II lines and a reduced dispersion in the abundances of other species compared to either the \citeauthor{2000ARep...44..611K} or the \citeauthor{2014A&A...563A.139P} models, and we conclude that our adopted model atmosphere parameters provide the best description of the atmosphere of RZ\,Psc at the epochs of our observations.
Our model yields an overall iron abundance [Fe/H]~=~--0.1 for RZ\,Psc.

Using the adopted model and the average Li EW of 207 m\AA\, we obtain a lithium abundance of A(Li)~=~3.05\footnote{A(x) = 12 + log n(x)/n(H), where n(x) is the number density of atoms of species x and n(H) is the number density of H atoms.} for RZ\,Psc.
This analysis uses the Li I line list from \citet{2002MNRAS.335.1005R}, including a six-component blend of the various explicit features that make up the  $^{7}$Li~+~$^{6}$Li feature, and assumes a standard isotopic ratio of 10/1.

\subsection{Spectral Type}
\label{sec:spectype}

Spectral types ranging from early K to late G have been assigned to RZ\,Psc in the literature thus far \citep[e.g.,][]{1960ApJ...131..632H,2000ARep...44..611K}.
It appears that RZ\,Psc is not a main sequence star, given optical spectral diagnostics that are indicative of low surface gravity \citep[][\S~\ref{sec:Atmosphere}]{2000ARep...44..611K}.
Our determination of an effective temperature of 5600~K and surface gravity of $\log{g}$~=~4.35~$\pm$~0.10 is consistent with a luminosity class of type III/IV.
The latter implies that RZ\,Psc must be either a pre-main sequence star or an evolved star (giant). 
Adopting this temperature and the empirical spectral type-temperature sequence for 5-30 Myr old pre-main sequence stars presented in \citet{2013ApJS..208....9P}, RZ\,Psc appears to have a spectral type of G4 if it is in a pre-main sequence evolutionary stage, but as \citet{1960ApJ...131..632H} originally noted, the photospheric features of RZ\,Psc correspond to a later spectral type, K0 IV.

\subsection{Stellar Activity and Rotation}
\label{sec:activityrotation}

\begin{figure}[h!]
 \includegraphics[width=0.75\hsize]{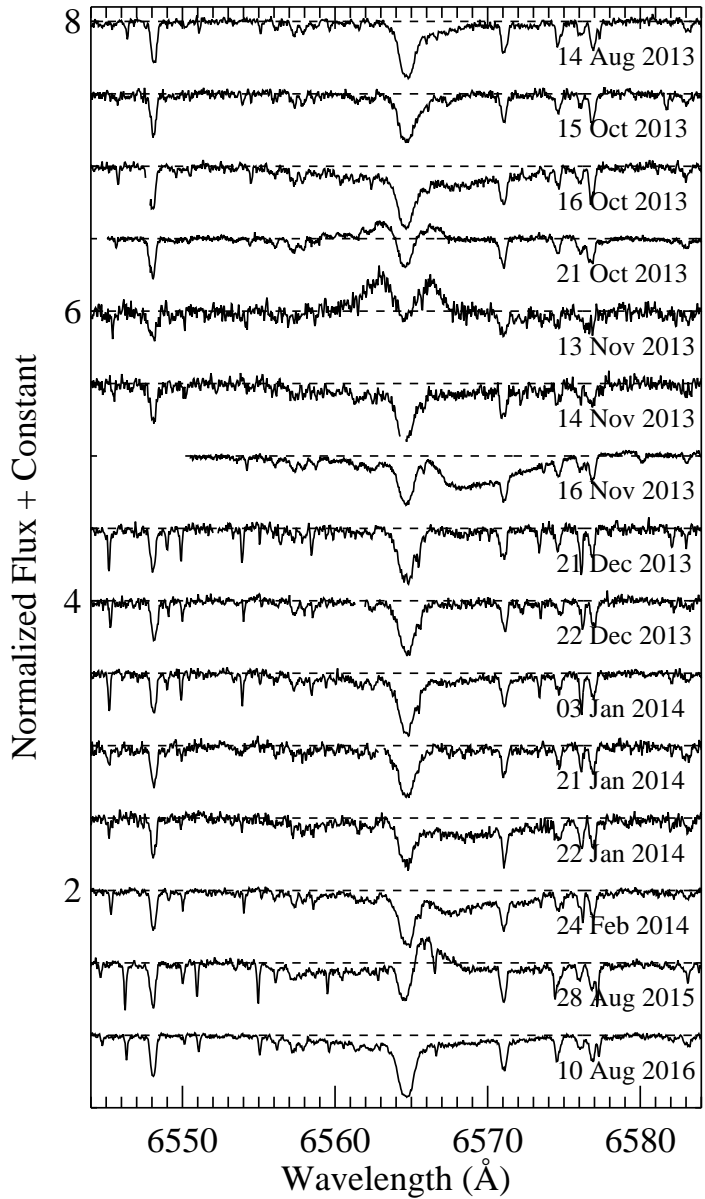}
\caption{\label{figrzhalpha} Keck HIRES and Lick Hamilton spectra of RZ\,Psc 
               showing the H$\alpha$ complex
               and surrounding region. All spectra are continuum-fit such that the continuum has a value
               of unity, then shifted vertically by a constant value for clarity. The dashed line for each
               epoch denotes the expected continuum value. 
               Wavelengths in this figure are plotted in
               vacuum. The wavelength scale for each 
               spectrum is shifted
               to match the 16 Nov 2013 epoch for clarity. Some cosmic rays and bad pixels have been
               removed manually from the spectra.}
\end{figure}

We examined Lick/Hamilton and Keck/HIRES optical spectra for RZ\,Psc for indications of stellar activity.
Typical features indicative of activity include H$\alpha$ emission and core-reversal emission in the Ca~{\sc ii} H, K, and infrared triplet lines.
Figure~\ref{figrzhalpha} showcases the dramatic variability displayed by RZ\,Psc in the H$\alpha$ region.
Variability in the strength of the photospheric absorption component of H$\alpha$ is evident, as is an additional, presumably circumstellar component that sometimes appears as broad absorption wings around the H$\alpha$ absorption line core \citep[e.g., 2013 August 14, 2013 October 16, and 2013 November 14; see also][]{2000ARep...44..611K} or, perhaps, as a stronger core absorption component (2013 November 16; all three spectra for the 2013 November 16 epoch have the same appearance to within their relative S/N).
Emission above the continuum level also appears at multiple epochs.

The highly variable H$\alpha$ line profile of RZ\,Psc, including the appearance of emission above the stellar continuum, is reminiscent of those of certain young stars that are rotating at near-breakup speeds and are possibly losing mass \citep[e.g., ][]{1985ApJ...288..259M,1989ApJ...346..160S}.
However, the v~sin~\textit{i} of RZ\,Psc (at most 23 km/s, see Table~\ref{tab:RZPscCharPrev}) is well below breakup speeds and is otherwise consistent with those of young cluster stars that display similar levels of X-ray activity \citep[e.g., ][their Figure 6]{1997ApJ...479..776S}.
This suggests a different origin for the circumstellar gas that is responsible for the variable H$\alpha$ emission from RZ\,Psc; we explore one potential explanation in Section~\ref{sec:origin}.

For each H$\alpha$ emission epoch we calculated the full-width of the double-peaked structures at 10\% of the peak line flux.
We find that the emission structures have velocity widths $\gtrsim$300\,km\,s$^{-1}$.
Thus, as was concluded by \citet{1989ApJ...346..160S} and \citet{1985ApJ...288..259M} for their rapidly-rotating stars, the H$\alpha$-emitting material is unlikely to be located at the stellar surface and instead is either being ejected in a wind or is orbiting or infalling within a few stellar radii of RZ\,Psc.

\begin{figure}[h!]
 \includegraphics[width=160mm]{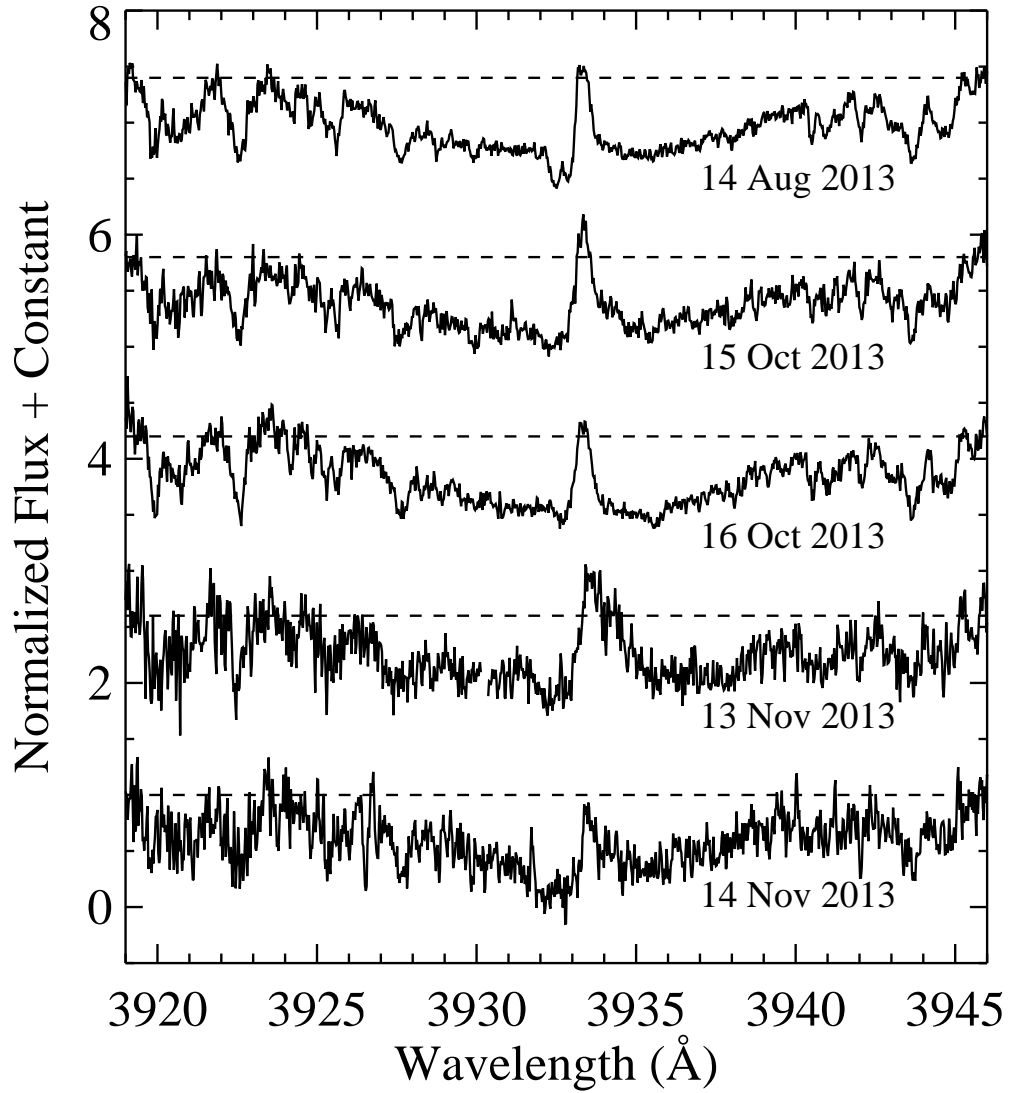}
\caption{\label{figrzcaiik} Lick Hamilton spectra of RZ\,Psc 
               showing the region around the Ca~II K line. The Ca~II H line is essentially identical in
               appearance for each epoch and thus not plotted here. 
               For each spectrum the blaze function is
               removed, then the spectrum is normalized to the flux level at $\approx$3946 \AA .
               The dashed line indicates this normalization level and can be used as a guide
               for changes in the core reversal emission level.
               Variability in the core reversal emission and blueshifted absorption components
               is evident.
               Wavelengths in this figure are plotted in
               air. The wavelength scale for each spectrum is shifted to match the 14 Aug 2013 epoch
               for clarity. Some cosmic rays have been manually removed from the spectra.}
\end{figure}

\begin{figure}[h!]
 \includegraphics[width=160mm]{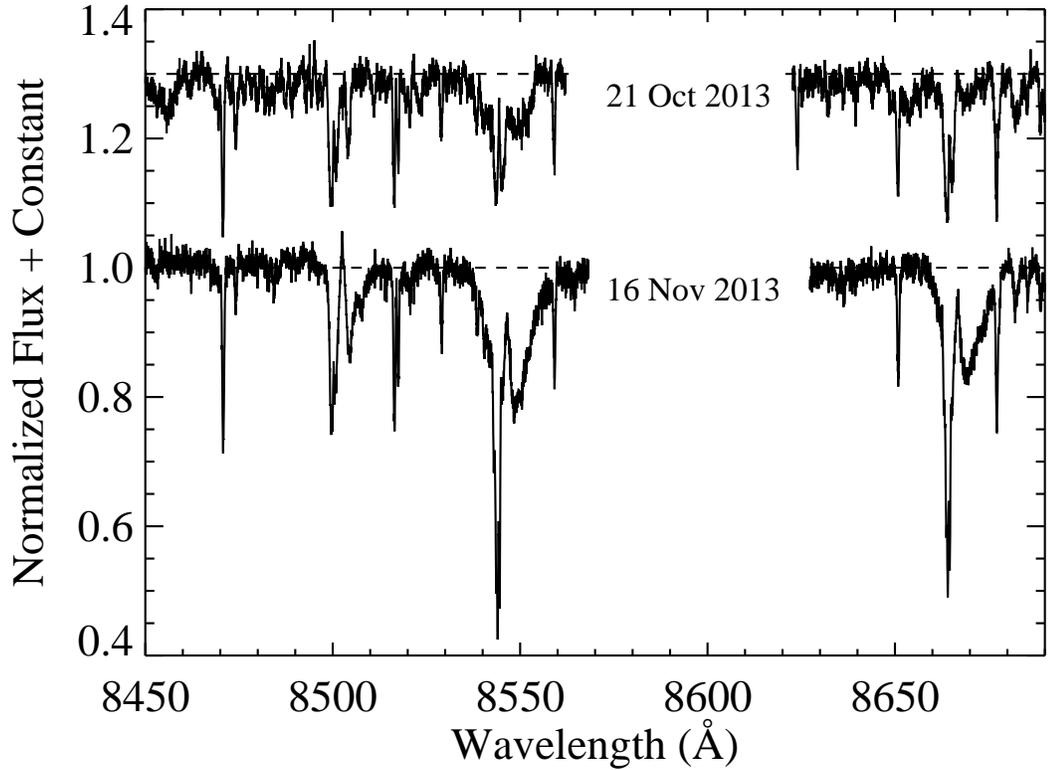}
\caption{\label{figrzcaiiirt} Keck HIRES spectra of RZ\,Psc 
               showing the Ca~II infrared triplet complex.
               Normalization and vertical shifting of spectra is performed as described in the caption
               to Figure~\ref{figrzhalpha}. Line variability is
               evident, with broad absorption components and core reversal emission.
               Spectra from 16 Nov 2013 taken later in the night are consistent with what is shown
               here to within their respective S/N.
               Wavelengths in this figure are plotted in
               vacuum. The wavelength scale of the spectra are not shifted in this figure. The lack of data 
               around 8600 \AA\ is due to the gap between red orders for the HIRES setup used.}
\end{figure}

Steadier emission likely associated with stellar activity is seen in the Ca~{\sc ii} H and K lines (Figure~\ref{figrzcaiik}).
In some epochs, a blueshifted absorption component is seen adjacent to the emission line, while the UT 2013 November 13 epoch shows a much broader emission line than is seen in other epochs.
It should be noted that the spectrum of UT 2013 November 13 is particularly noisy, perhaps as a result of RZ\,Psc being observed during during one of its deep optical minima.
The enhanced H$\alpha$ emission observed during this observation would be consistent with the star being heavily obscured by dust; such behavior is observed, e.g., in the case of the highly inclined for T Cha star/disk system \citep{2009A&A...501.1013S}.
Ca~II infrared triplet core-reversal emission is evident in the UT 2013 21 October 21 epoch and some emission is present in the UT 2013 November 16 epoch (Figure~\ref{figrzcaiiirt}), although its nature is not immediately obvious given the complex structure in the lines.
The high velocity of the emitting material intermittently seen in H$\alpha$ provides evidence for a wind from (or infall onto) the surface of the star.
A similar behavior was observed by \citet{2017A&A...599A..60P}.

\begin{figure}[h!]
 \includegraphics[width=160mm]{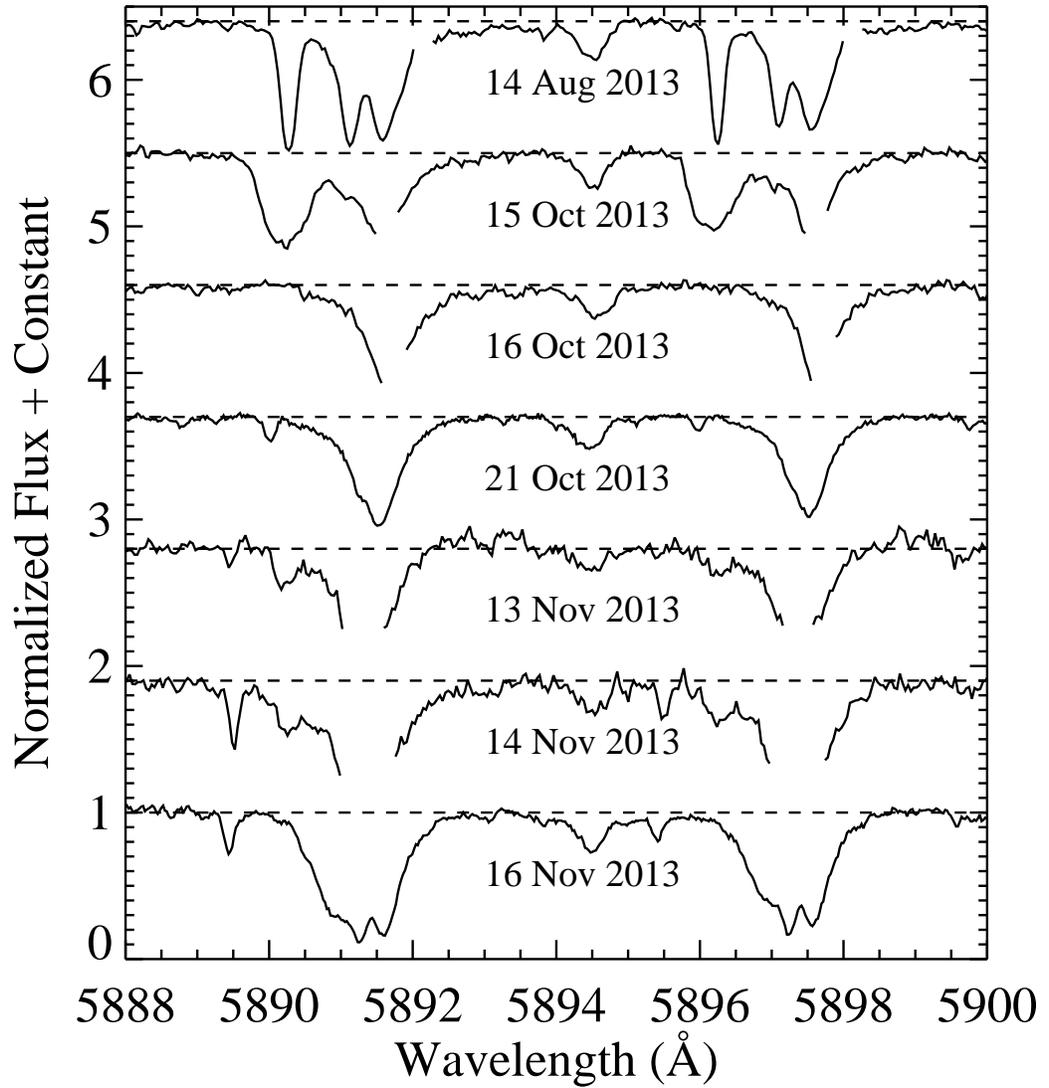}
\caption{\label{figrznad} Keck HIRES and Lick Hamilton spectra of RZ\,Psc 
               showing the Na~D doublet complex.
               Normalization and vertical shifting of spectra is performed as described in the caption
               to Figure~\ref{figrzhalpha}. 
               Wavelengths in this figure are plotted in
               vacuum. The wavelength scale for each 
               spectrum is shifted
               to match the 16 Nov 2013 epoch for clarity. In each Hamilton spectrum Na~D telluric 
               emission from nearby San Jose,
               CA is removed.}
\end{figure}

Figure~\ref{figrznad} displays the spectral region around the Na {\sc i} D lines.
Clear variability is evident, with blueshifted absorption features regularly appearing and moving away from the photospheric absorption components.
Such behavior was noted by \citet{2015AstL...41..407G}, and served as the basis for their hypothesis that RZ\,Psc is in the so-called ``propeller'' regime of mass loss due to a combination of strong magnetic field and low accretion rate.

\subsection{Radial Velocities}
\label{sec:RVAnalysis}

In order to derive RZ\,Psc's radial velocity (RV) at each epoch, we cross-correlated each of the HIRES and Hamilton spectra with spectra of stars with known radial velocities from \citet{2002ApJS..141..503N}.
This process also allowed us to search for any radial velocity variability between epochs.
Each of the HIRES and Hamilton spectra yields radial velocity measurements for RZ\,Psc with precision of roughly 0.2~km~s$^{-1}$ (Table~\ref{tabrvs}).
Excluding the UT 2013 November 13 epoch, there appears to be significantly detected radial velocity variability at the few km~s$^{-1}$ level.
This variability appears to be centered on a systemic velocity of roughly $-$1~km~s$^{-1}$, consistent with the radial velocity measurement determined by \citealt{2014Ap.....57..491P}, and does not have any obvious periodicity.
Such measurements seem at odds with the radial velocity of $-$11.75~$\pm$~1.1~km~s$^{-1}$ measured  by \citet{2000ARep...44..611K}.

\begin{figure}[h!]
 \begin{center}
 \begin{minipage}[!t]{60mm}
  \includegraphics[width=60mm]{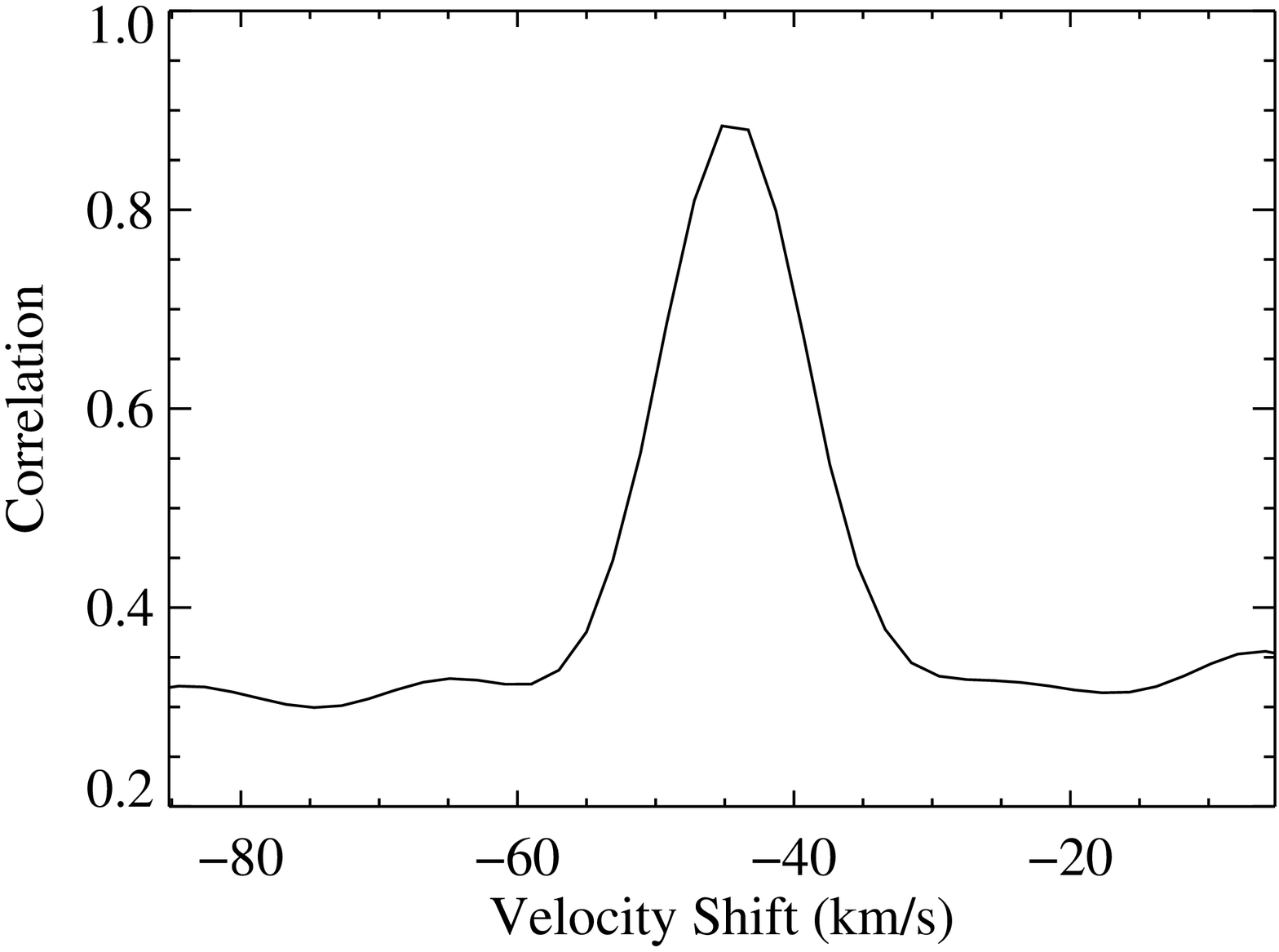}
 \end{minipage} 
 \\*
  \begin{minipage}[!h]{60mm}
  \includegraphics[width=60mm]{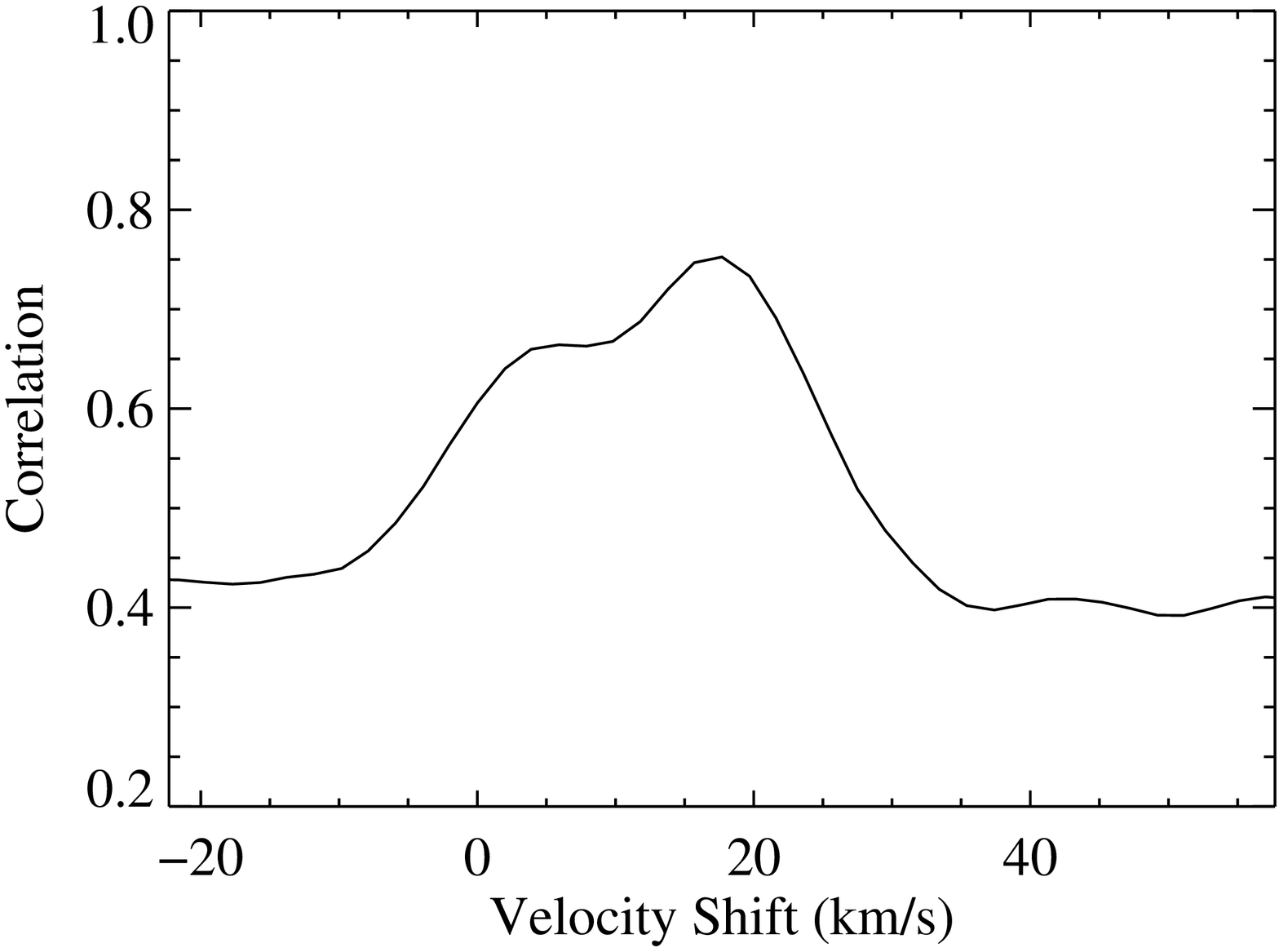}
 \end{minipage}
 \\*
 \begin{minipage}[!b]{60mm}
  \includegraphics[width=60mm]{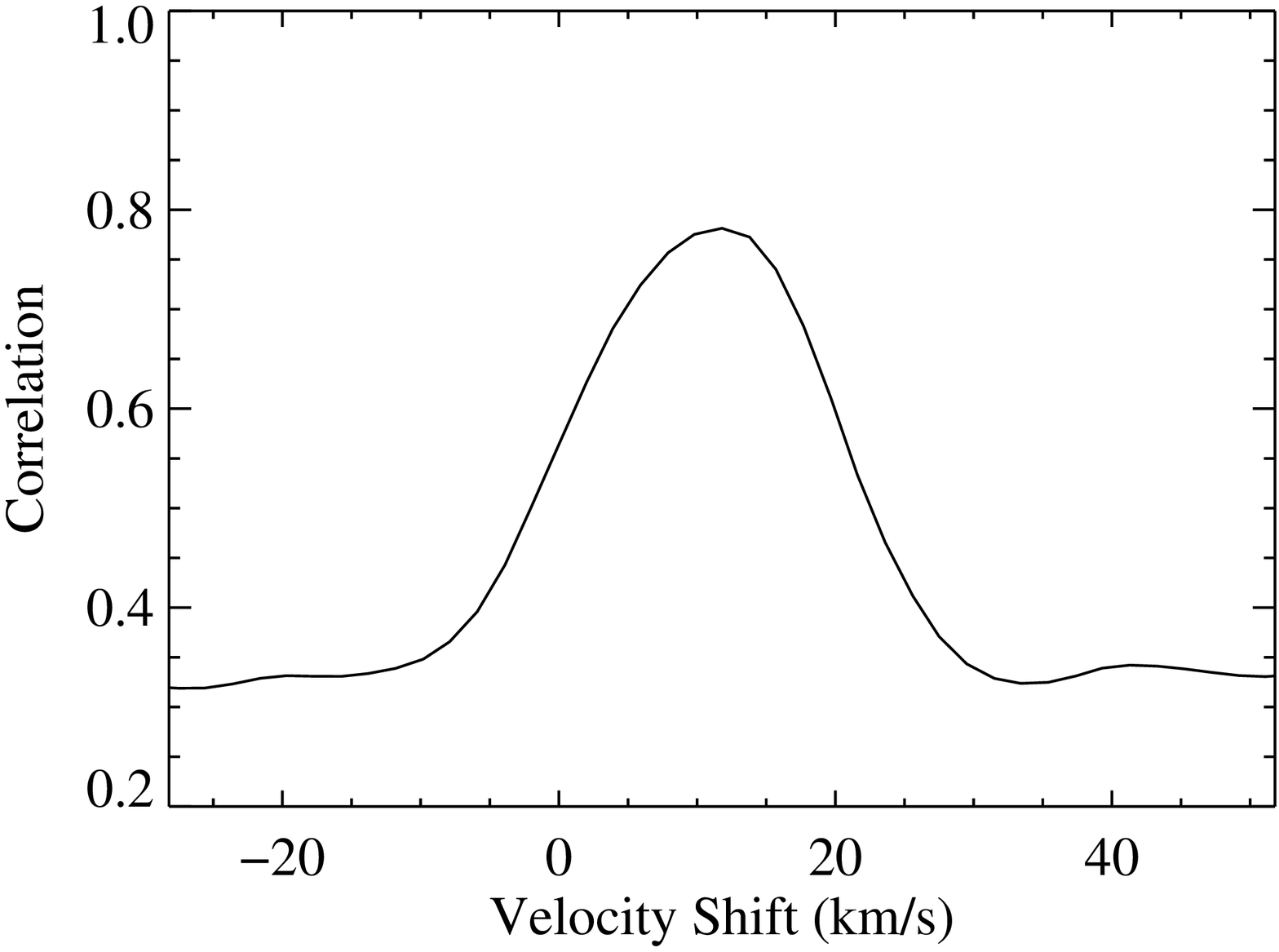}
 \end{minipage}
 \end{center}
\caption{\label{figrzxcor} Cross-correlation spectra between Lick Hamilton observations of RZ\,Psc 
               and radial velocity standard stars chosen from \citet{2002ApJS..141..503N}. Heliocentric velocity
               corrections have not been applied to any of the spectra. Wavelengths between 4800
               and 4857\,\AA\ are used for the cross-correlation in 
               each panel. {\it Top panel:} Cross-correlation
               between the bright standards HR\,124 and HR\,4027 using spectra from the
               night of UT 13 Nov 2013. A single, narrow
               cross-correlation peak is evident. {\it Middle panel:} Cross-correlation between RZ\,Psc
               and HR\,4027 using spectra from the night of UT 13 Nov 2013. A double-peaked
               cross-correlation peak is evident. Similar cross-correlation peak shapes are recovered
               when using different echelle orders. {\it Bottom panel:} Cross-correlation between RZ\,Psc
               and HR\,4027 using spectra from the night of UT 14 Nov 2013. A single cross-correlation
               peak is recovered, a feature seen in every order of these two spectra from this night
               and in general when performing
               cross-correlations for RZ\,Psc on nights other than UT 13 Nov 2013.}
\end{figure}

The velocity we determine from the Hamilton spectrum obtained on UT 2013 November 13 appears as a notable outlier when compared to almost all previous radial velocities measured for RZ\,Psc, apart from the \citet{2000ARep...44..611K} measurement.
At this epoch (UT 2013 November 13) the cross-correlation peak splits into two components (Figure~\ref{figrzxcor}; i.e., a redshifted absorption line and a weaker blueshifted component).
Indeed, when the UT 2013 November 13 spectrum is smoothed, a similar morphology is evident in some absorption lines that is not present in the UT 2013 November 14 spectrum.
A deblending fit to the cross-correlation double-peak was performed using the {\sf FXCOR} task within IRAF, and the resulting velocities for each peak (corrected to the heliocentric reference frame) are reported in Table~\ref{tabrvs}.
The unusually large negative velocity of the blue peak of the cross-correlation function for 2013 November 13 is marginally consistent with the velocity for RZ\,Psc determined by \citet{2000ARep...44..611K}\footnote{Unfortunately, it is not clear at which of these epochs (1991 December or 1998 August) \citet{2000ARep...44..611K} obtained this discrepant radial velocity measurement.}.

\subsection{Modeling the X-ray Spectrum of RZ\,Psc}
\label{sec:XrayModeling}

\begin{figure}[h!]
  \centering
  \includegraphics[width=0.6\hsize]{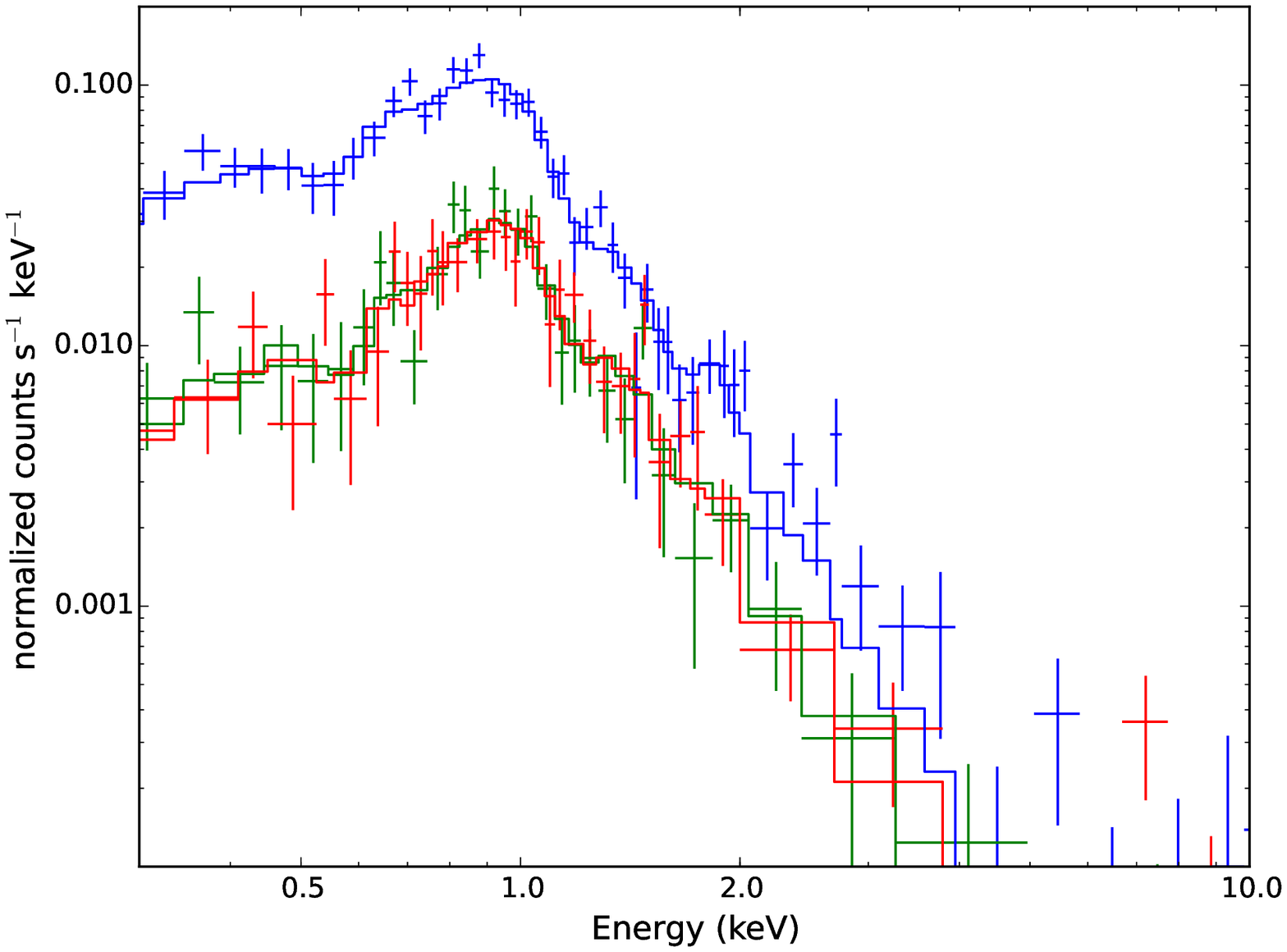}
   \includegraphics[width=0.6\hsize]{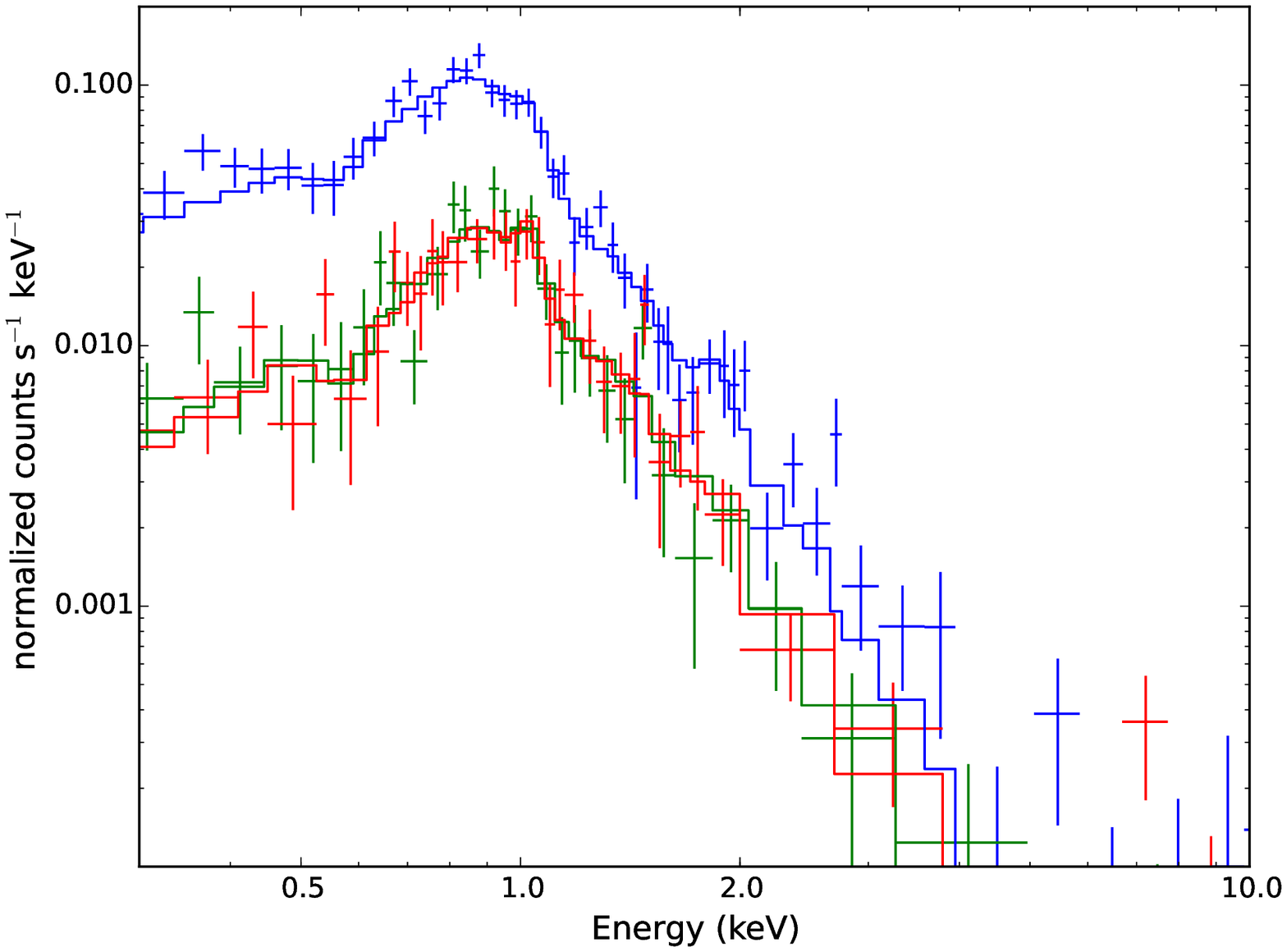}
    \includegraphics[width=0.6\hsize]{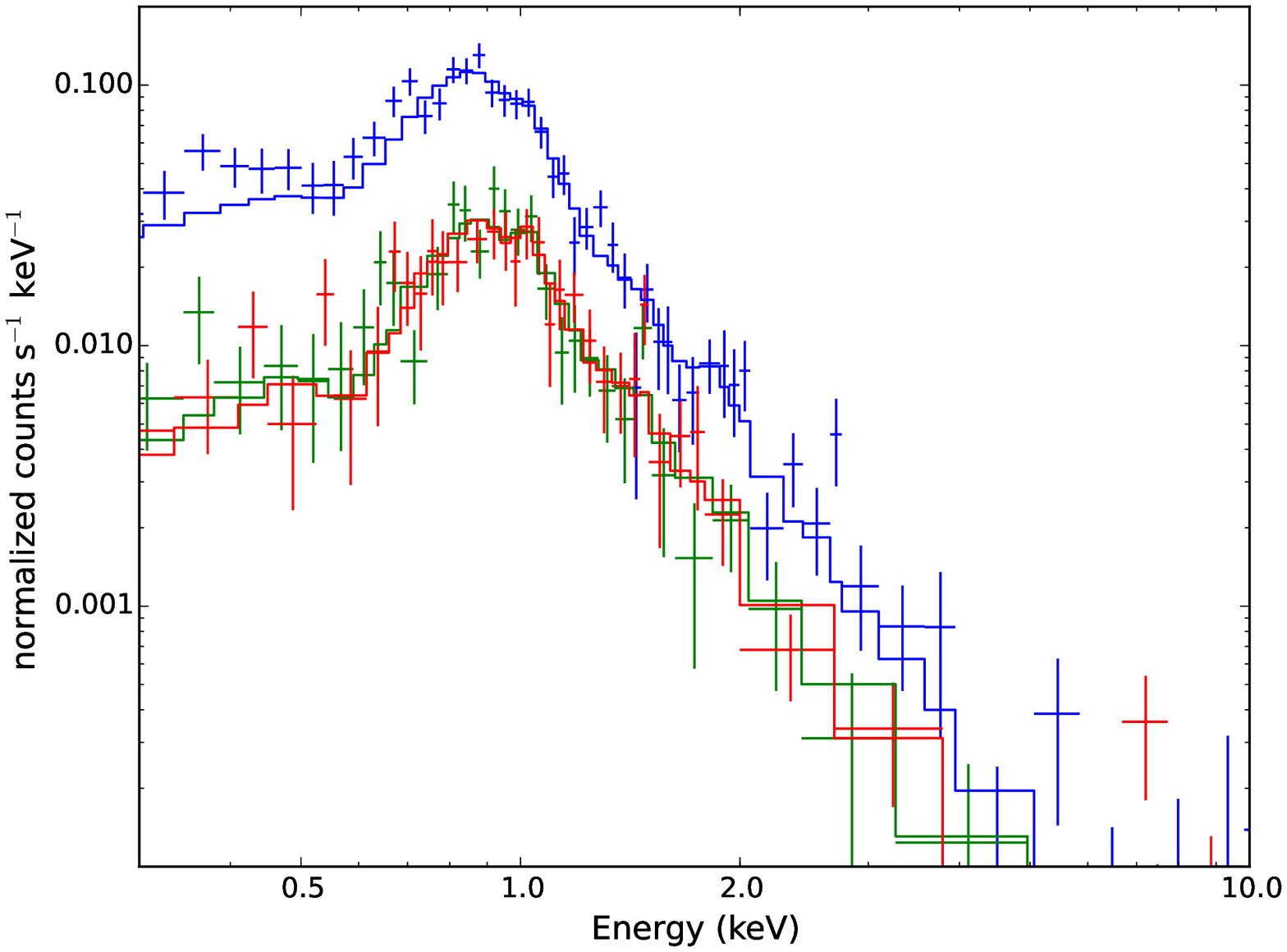}
  \caption{\textit{XMM-Newton} EPIC extracted spectra (crosses) of RZ\,Psc for pn (blue) and MOS (red and green) detectors. Overplotted are the best-fit models (histograms): Free Abundance Model (top), T Tauri Star Model (middle), Evolved Giant Star (bottom).}
  \label{fig:XRaySpectra}
\end{figure}

Source X-ray spectra were extracted from the \textit{XMM} pn, MOS1, and MOS2 event lists by  selecting photon events within circular regions with diameters of $\sim$ 25-40'' centered on RZ\,Psc (Figure~\ref{fig:XRayIR}).
Associated background spectra were extracted within circular regions with the same diameters from nearby, source-free regions.
We fit the resulting background-subtracted spectra of RZ\,Psc with the HEASOFT \textit{Xanadu}\footnote{See http://heasarc.gsfc.nasa.gov/docs/xanadu/xanadu.html.} software package (version 6.16), using XSPEC\footnote{See http://heasarc.gsfc.nasa.gov/xanadu/xspec.} version 12.8.2.
The pn, MOS1, and MOS2 X-ray spectra of RZ\,Psc, displayed in Figure~\ref{fig:XRaySpectra}, were analyzed using three different models over the range 0.15-10\,keV.
Each of the adopted models (see below) made use of the XSPEC optically thin thermal plasma model vapec \citep{2012ApJ...756..128F}\footnote{See http://www.atomdb.org.}, which is parameterized by plasma elemental abundances, temperature, and emission measure (indirectly, through the model normalization).
We included the potential effects of intervening absorption by using XSPEC's wabs absorption model \citep{1983ApJ...270..119M}.
All of our models include two temperature components, as required by the $\chi^{2}$ statistics.
The fixed and free parameters for all three models are displayed in Table~\ref{tbl:Models}, with the free parameters having associated uncertainty values.

Since the evolutionary status of RZ\,Psc was unclear to us, we adopted models that were representative of young stars and evolved stars.
The first model, hereafter referred to as ``T Tauri Star,'' uses plasma abundance parameter values that represent average values for T Tauri stars in Taurus (\citealt{2013ApJ...765....3S}, and references therein); these values are stated in Table~\ref{tbl:Models}.
For the second model, hereafter referred to as ``Evolved Giant Star,'' we fixed the parameters for the plasma abundances to values that have been determined to be typical for chromospherically active late-type giants \citep{2003A&A...409..263G,2003A&A...404..355G,2003A&A...400..249G}.
In the third model, designated ``Free Abundance,'' we allowed the abundances of Ne, Fe, and C to be free parameters.
Lines of the first two elements likely dominate the emission around the spectrum peak ($\sim$0.3-3 keV) in the relevant ($\sim$10 MK) temperature regime \citep[see, e.g.,][]{2002ApJ...567..434K}.
We allowed the abundance of C to be a free parameter in order to fit the spectral feature near 0.3-0.4\,keV.

The results of the spectral analysis are presented in Table~\ref{tbl:XRaySpecAnalysis} and Figure~\ref{fig:XRaySpectra}.
Adopting $D=174$ pc (see \S~\ref{sec:age}), we find a range of X-ray luminosities, $L_{X}$,  from $\sim 5.9 \times 10^{29}$ to $\sim1.5 \times 10^{30}$ ergs s$^{-1}$, depending on the model adopted.
The T Tauri and Evolved Giant models yield best-fit values of $L_{X}$ that are very similar and lie at the lower end of this range.
The corresponding range of best fit $N_{H}$ values are as small as $\sim4 \times 10^{20}$ cm$^{2}$ and as large as $\sim2 \times 10^{21}$ cm$^{2}$. 
The fit results for all three models indicate that the emission measures are similar for both temperature components, with the low- and high-temperature components lying in the ranges $\sim$$3-9$ MK and $\sim$$10-20$ MK, respectively.

\subsection{\textit{XMM-Newton} Optical Monitor Data}
\label{sec:OpticalData}

The \textit{XMM} Optical Monitor (OM) photometry was used to determine UBV and UVM2 magnitudes of RZ\,Psc simultaneously with the X-ray measurements.
These OM measurements thereby allow us to determine the star's bolometric flux $F_{\rm bol}$ during the X-ray observations, alleviating uncertainties in determining $L_{X}/L_{\rm bol}$ that could arise due to the significant optical variability and likely X-ray variability of RZ\,Psc.
The OM photometry is provided in Table~\ref{tbl:OMObs}.
It is evident that the star's flux remained constant in all four passbands during the observations.
Based on $V$-band magnitude obtained with the OM we further conclude that, during our \textit{XMM} observations, RZ\,Psc was not undergoing one of its rare, deep visual minima, which can be as dim as $V$ $\sim$ 14 \citep{2013A&A...553L...1D}.
Adopting the mean $V$-band magnitude obtained from the OM photometry and assuming bolometric corrections based on the adopted effective temperature and luminosity class of RZ\,Psc (see Section~\ref{sec:spectype}), we obtain $F_{\rm bol}$ values that range from $\sim 6.7 \times 10^{-10}$ to $\sim 6.9 \times 10^{-10}$ erg s$^{-1}$ cm$^{-2}$.
Adopting a distance of 170 pc to RZ\,Psc (see \S~\ref{sec:age}), we obtain a bolometric luminosity $L_{\rm bol}$ $\sim$ 0.6 L$_{\odot}$ (excluding any contribution from circumstellar dust).

From the OM photometric results and the adopted effective temperature, which implies $(B-V)_{0}$ = 0.66, we estimate the optical extinction of this system at the time of our {\it XMM-Newton} observations to be $A_{V} \sim 0.66$ \citep[assuming $R=3.09$;][]{1985ApJ...288..618R}.
Our estimated value of $A_{V}$ is larger than the value of $0.3~\pm~0.05$ quoted in \citet{2017A&A...599A..60P}.
The discrepancy between these values of $A_{V}$ are consistent with the optical dropout behavior associated with RZ\,Psc.

\subsection{Search for Comoving Low-Mass Stellar Companions}
\label{sec:companion}

The high galactic latitude of RZ\,Psc (see Table~\ref{tab:RZPscCharPrev}) and its large displacement from any known nearby young moving group \citep[see Table 1 of][]{2016IAUS..314...21M} are problematic for a young star scenario.
Thus, we used the pipeline-generated X-ray source list compiled from the \textit{XMM-Newton} observation as the basis for a search for young, X-ray active stars that are comoving with RZ\,Psc.

The \textit{XMM-Newton} images were matched to images from the WIYN 0.9m telescope and the 2MASS and \textit{WISE} image archives to identify optical and IR sources coincident with (i.e., within 2$”$ of) those XMM X-ray sources lying within approximately 15$'$ of RZ Psc.
This yielded seven X-ray sources with optical/IR counterparts, which we regarded as candidate comoving stars.
Spectral Energy Distributions (SEDs) for these seven sources were produced using magnitudes obtained from the WIYN 0.9m images, along with those available in the 2MASS and \textit{WISE} catalogs.
The resulting SEDs were then fit with stellar atmosphere models\footnote{The stellar atmosphere models were obtained from the Castelli and Kurucz grids of model atmospheres: http://kurucz.harvard.edu/grids.html.} \citep{2003IAUS..210P.A20C}.
This SED analysis and comparison of the proper motions of these seven candidates with that of RZ\,Psc eliminated all but one of these objects as possible members of a nearby moving group associated with RZ\,Psc.
This object, J0109+2758, is an M-type star at a separation of 2.3$'$ from RZ\,Psc \citep{2017AAS...22915422G}.
J0109+2578 was only detected by \textit{XMM}, and hence considered as a candidate comoving star, due to a strong flare during the exposure.
\citet{2017AAS...22915422G} concluded that J0109+2578 was a main sequence M flare star which, if associated with RZ\,Psc, would provide evidence in favor of an evolved star status.

To confirm the spectral type and evolutionary status of J0109+2578, we obtained a spectrum with the ESI on Keck~II.
Comparison with spectral templates in \citet{1984ApJS...56..257J} confirms that J0109+2578 is an M dwarf.
The Keck/ESI spectrum of J0109+2758 was cross-correlated with a spectrum of NLTT 853 \citep{2002ApJS..141..503N} that was obtained on the same night as J0109+2758 with the same setup.
From the cross-correlation, we measure a heliocentric radial velocity for J0109+2758 of $-$33.5 $\pm$ 1.5\,km\,s$^{-1}$, clearly discrepant with the approximate median radial velocity of RZ\,Psc ($V_{r}$ $\sim$ $-$2 km\,s$^{-1}$, see Table~\ref{tabrvs}).
Thus, we conclude that J0109+2758 is not comoving with RZ\,Psc and hence is not associated with it.

\section{Discussion}
\label{sec:discussion}

\subsection{Age Diagnostics: Stellar Activity and Li Line Strength}
\label{sec:age}

Our {\it XMM-Newton} observations reveal that RZ\,Psc is highly X-ray active, with
 $\log{(L_X/L_{\rm bol})}$ in the range $-3.7$ to $-3.2$ (\S~\ref{sec:XrayModeling}).
This large fractional X-ray luminosity is consistent with the enhanced presence of activity indicators in the optical spectrum of RZ\,Psc (\S~\ref{sec:activityrotation}).
Such levels of X-ray activity are primarily associated with late-type, short-period binaries (e.g., RS\,CVn systems) or rapidly rotating, late-type single stars \citep[][and references therein]{2005LRSP....2....8B, 2005A&A...444..531G}.
Stars in the latter category are generally either young --- i.e., in the T Tauri, post-T Tauri, or zero-age main sequence stages --- or are particularly fast-rotating evolved giants (e.g., stars of the FK Com class).
However, whereas virtually all young stars with effective temperatures similar to RZ\,Psc (i.e., G type stars) have values of $L_{X}/L_{\rm bol}$ in the range we determine for RZ\,Psc \citep[e.g.,][see their Figure 6]{1995A&A...300..134R}, very few rapidly rotating giants have such large relative X-ray luminosities \citep{2005A&A...444..531G}.
Thus, in contrast to the case of BP\,Psc \citep{2010ApJ...719L..65K}, the value of $\log{(L_X/L_{\rm bol})}$ obtained from our X-ray observations strongly favors young-star status for RZ\,Psc. 

Our measurement of a large value of $\log{(L_X/L_{\rm bol})}$, which is indicative of strong surface magnetic activity, supports the hypothesis that RZ\,Psc is in the ``magnetic propeller'' regime of efficient mass loss relative to its accretion rate \citep{2015AstL...41..407G}.
To attempt to better determine the nature of the X-ray emission from RZ\,Psc, it would be beneficial to obtain high-resolution (gratings) X-ray spectra of the star, from which we can obtain the detailed emission-line spectrum of the X-ray emitting plasma.
Such data might establish, e.g., whether the composition of this plasma resembles those characteristic of the X-ray-emitting plasma of pre-main sequence stars and whether some of the X-ray emission from RZ\,Psc may arise from accretion activity \citep[e.g.,][]{2010ApJ...710.1835B}.

Our measurement of the EW of the Li~I line, averaged over all epochs, is 207 $\pm$ 6\,m\AA, consistent with that measured by \citet{2010A&A...524A...8G}.
This EW(Li) measurement is somewhat less than those of 10-20 Myr old G-type stars (see Figure~6, \citealt{2014A&A...567A..55S}), but is consistent with G-type stars of ages 30-50 Myr (Table~2, \citealt{2001A&A...372..862R} and Figure~3, \citealt{2004ARA&A..42..685Z}).
The EW(Li) of RZ\,Psc is somewhat larger than that of G-type stars in the Pleiades \citep{1993AJ....106.1059S}, but consistent with such stars in the $\beta$ Pictoris Moving Group (age $\sim$ 23 $\pm$ 3 Myr; \citealt{2016IAUS..314...21M}) and the Tucana-Horologium Association (age $\sim$ 45 $\pm$ 5 Myr; \citealt{2016IAUS..314...21M}) \citep[see Figure~5;][]{2008ApJ...689.1127M}.

Comparing with Table~1 in \citet{1993AJ....106.1059S}, we estimate A(Li) $\sim$ 3.0 for RZ\,Psc, consistent with the value of 3.05 that we calculated in $\sec$~\ref{sec:Atmosphere}.
Although there are approximately 40 examples of Li-rich giants with such large lithium abundances \citep[e.g.,][]{2016MNRAS.461.3336C}, almost all of these giants have much smaller values of log g (log g $<$ 3) than RZ\,Psc (log g $\sim$ 4.35 $\pm$ 0.1).

\begin{figure}[h!]
  \centering
  \includegraphics[width=1.0\hsize]{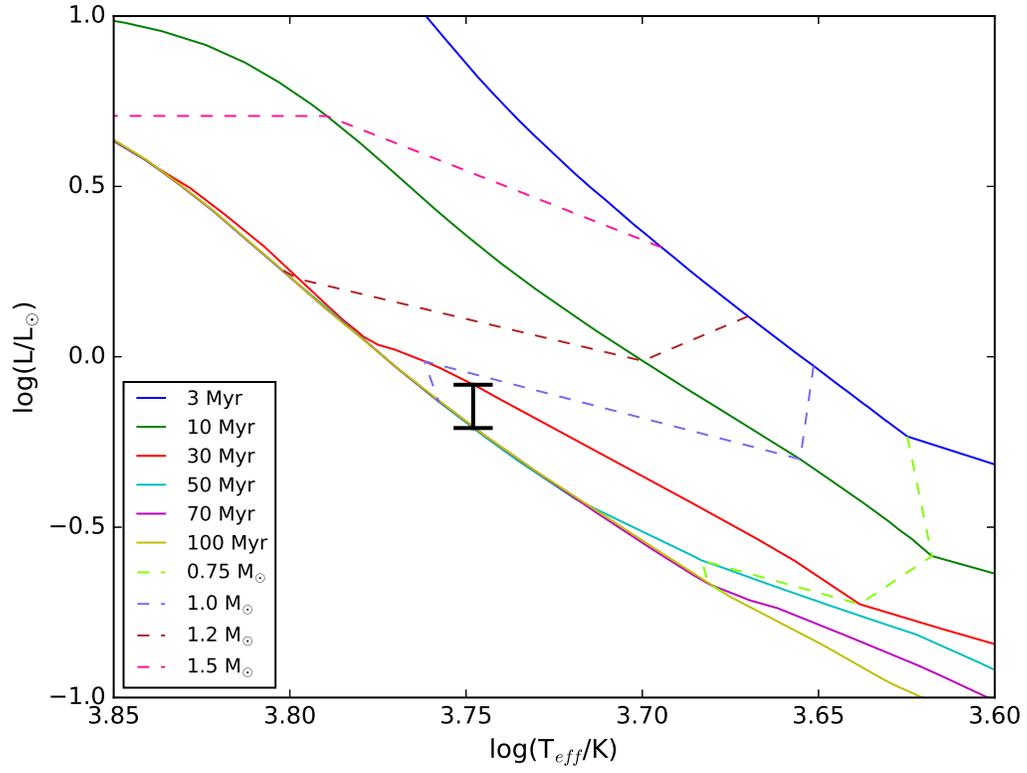}
  \caption{The H-R diagram position of RZ\,Psc, assuming an age range of 30-50 Myr, as shown by the vertical line, overlaid with PARSEC pre-main sequence isochrones and pre-main sequence tracks for a variety of masses \citep{2012MNRAS.427..127B}.}
  \label{fig:isochrones}
\end{figure}

In light of this evidence for pre-main sequence status, we have conducted an H-R diagram analysis with theoretical pre-main sequence isochrones to estimate the likely range of luminosity of RZ\,Psc.
Following our Li EW analysis, we assume an age range of 30-50 Myr for RZ\,Psc.
The resulting H-R diagram is displayed in Figure~\ref{fig:isochrones}, where the isochrones and pre-main sequence tracks overlaid are obtained from PARSEC\footnote{http://stev.oapd.inaf.it/cgi-bin/cmd} \citep[the PAdova and TRieste Stellar Evolution Code;][]{2012MNRAS.427..127B}.
Figure~\ref{fig:isochrones} suggests that RZ\,Psc is a pre-main sequence star of mass 0.75-1.0 M$_{\odot}$.
Our H-R diagram analysis yields $\log{(L_{*}/L_{\rm bol})}$ in the range of $-0.08$ to $-0.21$, which corresponds to a distance to RZ\,Psc of $\sim$ 186 pc if RZ\,Psc is 30 Myr old and a distance of $\sim$ 161 pc if RZ\,Psc is 50 Myr old.
This estimated distance range, obtained from our Li- and X-ray-based age range, agrees with the distance estimate of 160 pc mentioned in \citet[][and references therein]{2017A&A...599A..60P}.
Our age and mass estimates are also similar to the values cited by those authors (25 $\pm$ 5 Myr and 1 M$_{\odot}$, respectively).

Finally, we consider the \textit{UVW} velocities of RZ\,Psc.
Adopting a median estimated distance of 170 pc, the Table~\ref{tab:RZPscCharPrev} proper motion, and a median radial velocity of $-$2 km s$^{-1}$ (Table~\ref{tabrvs}), the \textit{UVW} are $-$13.4, $-$18.2, $-$5.4 km s$^{-1}$.\footnote{http://kinematics.bdnyc.org/query}
These \textit{UVW} values are consistent with young star status -- the closest kinematic matches being to the TW Hya Association, $\beta$ Pic moving group, and the Columba Association \citep{2016IAUS..314...21M}.
However, RZ\,Psc is much further from Earth than established members of these groups.

\subsection{X-ray Absorption versus Optical Extinction}

\begin{figure}[h!]
  \centering
  \includegraphics[width=1.0\hsize]{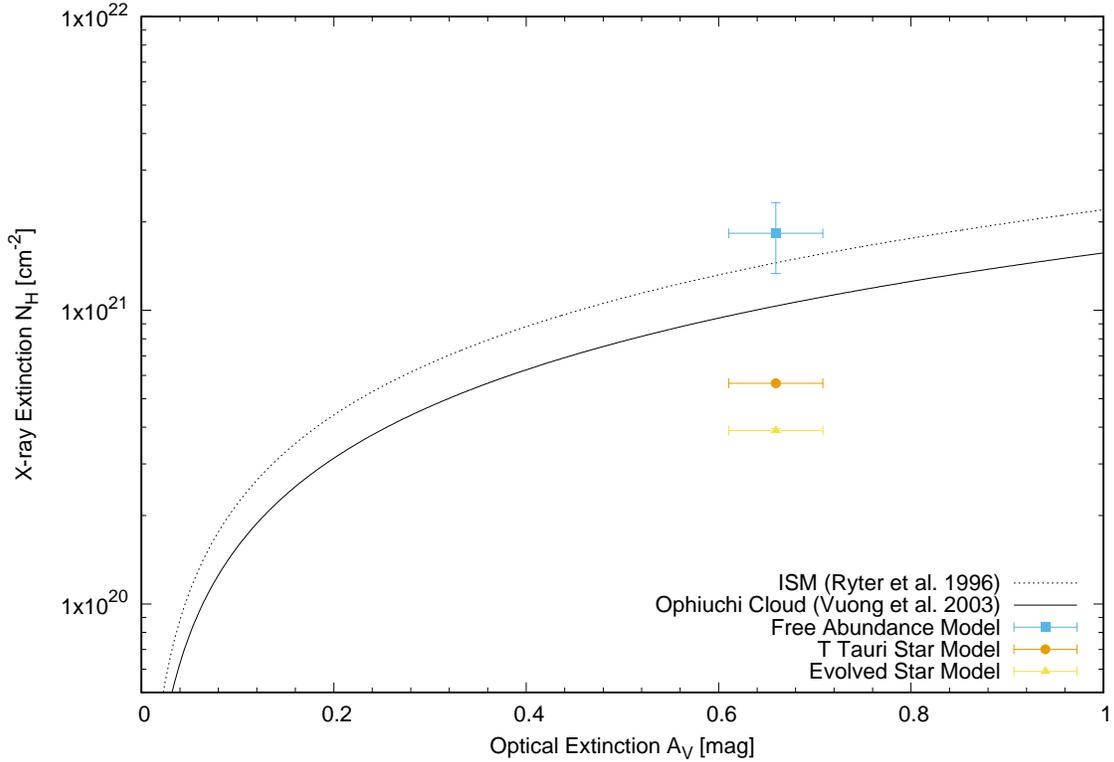}
  \caption{X-ray absorption column density $N_{H}$ versus visual extinction $A_{V}$ for RZ\,Psc compared to the local ISM (black dotted line; \citealt{1996Ap&SS.236..285R}) and the $\rho$ Ophiuchi molecular cloud (black solid line; \citealt{2003A&A...408..581V}), where we assume $R_{V}$ = 3.1. We plot the X-ray extinction values derived from spectral fitting for the derived spectral type of G4 and two different evolutionary stages (pre- and post-main sequence, respectively). The error bars represented are formal errors on the fit and do not reflect systematic uncertainties (i.e., the implicit assumption that we know all the abundances in the models).}
  \label{fig:NHAV}
\end{figure}

In Figure~\ref{fig:NHAV}, we compare the range of the gas column density we determine from the photoelectric absorption of the X-rays emitted by RZ\,Psc, $N_{H} \sim (0.4-1.8) \times 10^{21}$ cm$^{-2}$ (\S~\ref{sec:XrayModeling}) with the estimate of visual extinction due to dust ($A_{V}$) along the line of sight to the star afforded by the simultaneous {\it XMM-Newton} OM data (\S~\ref{sec:OpticalData}).

These measurements for RZ\,Psc are overlaid over empirical $N_{H}$ versus $A_{V}$ curves for the local ISM and the $\rho$ Ophiuchi molecular cloud.
The figure demonstrates that the relation between absorption and extinction toward RZ\,Psc is consistent with that of the ISM if we adopt the ``Free Abundance'' model; i.e.,  the model results in an $N_{H}$ that is most consistent with an $A_{V}$ given ISM-like grains.
Adopting either of the other X-ray spectral models would imply that the gas-to-dust ratio of the circumstellar material may differ significantly from that of the ISM or $\rho$ Oph.
Such an interpretation of Figure~\ref{fig:NHAV} is subject to the important caveat that, because the OM $V$ photometry indicates that RZ\,Psc was observed in a relatively low-extinction state, we cannot ascertain whether the observed relationship between $N_{H}$ and $A_{V}$ toward RZ\,Psc at the time of our observations was representative of the intervening ISM or of circumstellar material associated with RZ\,Psc itself.

\subsection{The Puzzling Radial Velocity Behavior of RZ\,Psc}
\label{sec:RV}

In our radial velocity (RV) cross-correlation analysis, we find marginal evidence for RV variability, including one epoch in which the cross-correlation function appears double-peaked (Figure~\ref{figrzxcor}).
One possibility that could account for this potential RV variability is that RZ\,Psc may reside in a spectroscopic binary system, one that can, on rare occasions, be double lined.
This would make RZ\,Psc superficially similar to BD+20 307, which also has a large value of $L_{\rm IR}/L_{\rm bol}$, thought to be the result of a catastrophic collision of terrestrial mass planets \citep[][]{2008ApJ...688.1345Z}.

If RZ\,Psc is indeed a spectroscopic binary, then there would be a new, interesting route through which we could explore the age and evolutionary status of the system.
If both stars are pre-main sequence, then both should have comparably strong lithium absorption.
We examined the line shape of the UT 2013 November 13 Li~I\,$\lambda$6708 line to see if it shows a blueshifted shoulder.
First, the EW of the Li~I line was measured in all epochs.
To within the uncertainties of each epoch, these EWs agreed.
Next, a smoothed Li~I line was compared to other comparably strong lines at similar wavelengths, most notably the Ca~I\,$\lambda$6717 line.
While the Ca~I line showed an obvious slightly weaker blueshifted shoulder, the Li~I line did not.
Finally, cross-correlations were performed between the UT 2013 November 13 spectrum of RZ\,Psc and other stars using only single lines.
Most single lines showed either double-peaked cross-correlation maximums or significant broadening in the blueshifted region where the second peak would be expected.
The Li I line, in contrast, was well fit with a single gaussian and showed no signs of blueshifted broadening.
The velocity of the Li I line, however, does not match either velocity measured on UT 2013 November 13 via cross-correlation analysis, but rather lies in between that of these two velocity signals.
This complication makes it difficult to interpret the Li I as present in only one star in a putative binary system.

\subsection{What is the Origin of the Circumstellar Material Orbiting RZ\,Psc?}
\label{sec:origin}

The extreme dropout events in the optical light curve of RZ\,Psc invite comparisons between this system and objects showing similarly dramatic dropouts, such as the variable field star KIC 8462852 \citep[aka ``Tabby's Star'';][]{2016MNRAS.457.3988B} and the remarkable ``polluted'' white dwarf WD 1145+017 \citep[e.g.,][]{2016MNRAS.458.3904R}.
The profound, seemingly aperiodic, variability observed in both of those cases has been cited as evidence for the presence of orbiting and/or infalling circumstellar debris arising either from a catastrophic collision or tidal stripping of a subplanetary-mass body or bodies.
\citet{2016ApJ...816L..22X} suggest that the atomic absorption lines observed in the spectrum of WD 1145+017 could come from either a burst of accretion due to disintegrating planetesimals, a previous tidal disruption, or both.

As in the case of WD 1145+017, which displays wide and variable gaseous absorption lines in its spectrum that are indicative of a gas-rich disk \citep{2016ApJ...816L..22X}, the presence of rapidly variable emission and absorption in the wings of RZ\,Psc's H$\alpha$ line profiles (Figure~\ref{figrzhalpha}) suggests that its orbiting debris includes a significant gaseous component.
However, in terms of evolutionary state and, hence, the nature of the disrupted orbiting body or bodies, the (G-type) RZ\,Psc system would appear to have more in common with (F-type) KIC 8462852 than WD 1145+017.
Indeed, a major difference between the RZ\,Psc and KIC 8462852 systems would appear to be that the circumstellar mass associated with RZ\,Psc is far larger than that associated with KIC 8462852.
Specifically, unlike RZ\,Psc, KIC 8462852 does not have a detectable IR excess, and its optical dropouts are far less pronounced; whereas RZ\,Psc suddenly dims by several magnitudes \citep{2013A&A...553L...1D}, the sudden dips in flux exhibited by KIC 8462852 are of the order of $\sim$20 percent or less \citep{2016MNRAS.457.3988B}.
 This suggests that the putative body (or bodies) destroyed around RZ\,Psc was far more massive than in the case of the KIC 8462852 system, whose debris has variously been attributed to a rocky body originally a few hundred km in diameter with a mass of at most $10^{-6}$ Earth masses \citep{2016MNRAS.457.3988B} or to a handful of disintegrating cometary bodies \citep{2017A&A...600A..86N}.

We hence propose that the puzzling variability behavior and enormous infrared excess of RZ\,Psc is most readily ascribed to the aftermath of the recent tidal disruption of a substellar companion or giant planet, or a catastrophic collision involving one or more relatively massive, gas-rich orbiting bodies.
Evidently, as in the cases of KIC 8462852 and WD 1145+017, the enormous dips in the optical light curve of RZ\,Psc require that the orbiting debris resulting from this destructive event is confined to a disk that lies nearly along our line of sight to the star.
Although the bulk of the dusty debris is likely orbiting $\sim$0.3 au from the star \citep[based on the temperature of the dust excess, $\sim$500 K;][]{2013A&A...553L...1D}, the broad absorption features in the H$\alpha$ line profiles of RZ\,Psc indicate that at least some of the circumstellar material is either accreting onto the star, outflowing, or both.

Although the preponderance of evidence appears to support the young star status of RZ\,Psc, there are some caveats to be considered.
The consumption of a giant planet or substellar companion could be polluting the atmosphere of RZ\,Psc, thereby increasing the atmospheric abundance of Li, which would make the star appear younger than it is \citep[see, e.g.,][]{2002ApJ...572.1012S}.
Our measured (relatively low) value of log g could similarly be explained by accretion of debris from a substellar-mass companion, given that the primary would be expected to expand during the accretion process.
Indeed, it is possible that log g may be variable as a consequence of this (presumably ongoing) planet or substellar companion consumption process.
Accretion of material from a disrupted massive body would also increase the magnetic activity of the star, and hence might also explain the prodigious X-ray output of RZ\,Psc.
Thus, we should look to \textit{Gaia}'s forthcoming determination of the parallax distance and refinement of the space velocity of RZ\,Psc as the prime means to verify its youth.

\section{Summary and Conclusions}
\label{sec:summary}

We have used \textit{XMM-Newton}, along with high-resolution optical spectroscopy, to characterize the properties of the infrared-excess, variable star RZ\,Psc so as to confirm its evolutionary status.
The \textit{XMM-Newton} observations produced a detection of a bright X-ray point source coincident with the centroids of optical and infrared emission at RZ\,Psc, with the log of the ratio of X-ray to bolometric luminosity, $\log{L_X/L_{\rm bol}}$, in the range $-3.7$ to $-3.2$.
These results are consistent with $\log{L_X/L_{\rm bol}}$ ratios typical of low-mass, pre-main sequence stars, and larger than that of all but the most X-ray-active stars among giants (e.g., FK Com-type giants).
Examination of the X-ray sources in the RZ\,Psc field yields one candidate comoving (M dwarf) star, but the radial velocity of this potential wide (2.3$'$ separation) companion is inconsistent with that of RZ\,Psc.

High-resolution optical spectra obtained with the Hamilton Echelle on the Lick Shane 3\,m telescope and HIRES on the Keck\,I 10\,m telescope indicate that RZ\,Psc has an effective temperature and a surface gravity that are consistent with a pre-main sequence star.
Sporatic radial velocity variability may also be observed in RZ\,Psc; if confirmed, this would suggest that it may be a spectroscopic binary system.
We note that the potential radial velocity variability and bizarre H$\alpha$ emission-line profile variability observed for RZ\,Psc are both reminiscent of the 5-10 Myr old star T Cha, which also exhibits deep absorption episodes, like RZ\,Psc, due to an inclined dusty disk \citep{2009A&A...501.1013S}.

The \textit{XMM-Newton} and high-resolution optical spectroscopy results favor a young star status for RZ\,Psc.
Measurements of the Li EW indicate that RZ\,Psc is a $\sim$ 30-50 Myr old post-T Tauri star.
If the age of RZ\,Psc is indeed this advanced, the presence of significant, varying column densities of circumstellar gas and dust renders it extremely unusual among Sun-like pre-main sequence stars.
By analogy with objects such as KIC 8462852 \citep[aka ``Tabby's Star'';][]{2016MNRAS.457.3988B} and WD 1145+017 \citep[e.g.,][]{2016MNRAS.458.3904R}, it is possible that, in the RZ\,Psc system, we are seeing evidence of a catastrophic event, for example, the destruction of a massive planet.
Optical spectral features indicative of activity and/or circumstellar material, such as core-reversal emission in the Ca~{\sc ii} H, K, and infrared triplet lines and H$\alpha$ emission, are present in our spectra over multiple epochs, and provide evidence for the presence of a significant mass of circumstellar gas associated with RZ\,Psc.
The presence of a significant mass of circumstellar gas (as reflected in the broad H$\alpha$ emission-line profiles) might imply that the cannibalized planet was a hot Jupiter.

An H-R diagram analysis indicates a distance to RZ\,Psc of $\sim$ 170 pc if RZ\,Psc is a pre-main sequence star.
\textit{Gaia} should provide the parallax distance and space velocity measurements necessary to nail down the evolutionary status of RZ\,Psc and to refine estimates of its age.
Further observations are warranted to understand the nature of this enigmatic star: high-resolution X-ray spectroscopy of RZ\,Psc would improve constraints on the abundances of RZ\,Psc's X-ray-emitting plasma; an optical and infrared spectroscopy campaign would shed light on the potential binary nature of the system; and submillimeter interferometric imaging and optical/IR coronographic adaptive optics imaging would establish whether there is cold, extended gas and dust associated with the RZ\,Psc disk.

\acknowledgments{\it This research was supported in part by NASA Astrophysics Data Analysis program grant NNX16AG13G to RIT.
Carl Melis acknowledges NASA Astrophysics Data Analysis program grant ADAP13-0178.
This paper is based in part on observations at Kitt Peak National Observatory, National Optical Astronomy Observatory (NOAO), which is operated by the Association of Universities for Research in Astronomy (AURA) under a cooperative agreement with the National Science Foundation.
Some of the data presented herein were obtained at the W. M. Keck Observatory, which is operated as a scientific partnership among the California Institute of Technology, the University of California and the National Aeronautics and Space Administration.
The Observatory was made possible by the generous financial support of the W. M. Keck Foundation.
The authors wish to recognize and acknowledge the very significant cultural role and reverence that the summit of Maunakea has always had within the indigenous Hawaiian community.
We are most fortunate to have the opportunity to conduct observations from this mountain.
We thank the referee Ilya Potravnov for an insightful and thorough review of our work.}

\bibliography{References}

\begin{thebibliography}{}
\expandafter\ifx\csname natexlab\endcsname\relax\def\natexlab#1{#1}\fi

\bibitem[{{Asplund} {et~al.}(2009){Asplund}, {Grevesse}, {Sauval}, \&
  {Scott}}]{2009ARA&A..47..481A}
{Asplund}, M., {Grevesse}, N., {Sauval}, A.~J., \& {Scott}, P. 2009, \araa, 47,
  481

\bibitem[{{Berdyugina}(2005)}]{2005LRSP....2....8B}
{Berdyugina}, S.~V. 2005, Living Reviews in Solar Physics, 2,
  doi:10.12942/lrsp-2005-8

\bibitem[{{Boyajian} {et~al.}(2016){Boyajian}, {LaCourse}, {Rappaport},
  {Fabrycky}, {Fischer}, {Gandolfi}, {Kennedy}, {Korhonen}, {Liu}, {Moor},
  {Olah}, {Vida}, {Wyatt}, {Best}, {Brewer}, {Ciesla}, {Cs{\'a}k}, {Deeg},
  {Dupuy}, {Handler}, {Heng}, {Howell}, {Ishikawa}, {Kov{\'a}cs}, {Kozakis},
  {Kriskovics}, {Lehtinen}, {Lintott}, {Lynn}, {Nespral}, {Nikbakhsh},
  {Schawinski}, {Schmitt}, {Smith}, {Szabo}, {Szabo}, {Viuho}, {Wang},
  {Weiksnar}, {Bosch}, {Connors}, {Goodman}, {Green}, {Hoekstra}, {Jebson},
  {Jek}, {Omohundro}, {Schwengeler}, \& {Szewczyk}}]{2016MNRAS.457.3988B}
{Boyajian}, T.~S., {LaCourse}, D.~M., {Rappaport}, S.~A., {et~al.} 2016,
  \mnras, 457, 3988

\bibitem[{{Bressan} {et~al.}(2012){Bressan}, {Marigo}, {Girardi}, {Salasnich},
  {Dal Cero}, {Rubele}, \& {Nanni}}]{2012MNRAS.427..127B}
{Bressan}, A., {Marigo}, P., {Girardi}, L., {et~al.} 2012, \mnras, 427, 127

\bibitem[{{Brickhouse} {et~al.}(2010){Brickhouse}, {Cranmer}, {Dupree}, {Luna},
  \& {Wolk}}]{2010ApJ...710.1835B}
{Brickhouse}, N.~S., {Cranmer}, S.~R., {Dupree}, A.~K., {Luna}, G.~J.~M., \&
  {Wolk}, S. 2010, \apj, 710, 1835

\bibitem[{{Casey} {et~al.}(2016){Casey}, {Ruchti}, {Masseron}, {Randich},
  {Gilmore}, {Lind}, {Kennedy}, {Koposov}, {Hourihane}, {Franciosini}, {Lewis},
  {Magrini}, {Morbidelli}, {Sacco}, {Worley}, {Feltzing}, {Jeffries},
  {Vallenari}, {Bensby}, {Bragaglia}, {Flaccomio}, {Francois}, {Korn},
  {Lanzafame}, {Pancino}, {Recio-Blanco}, {Smiljanic}, {Carraro}, {Costado},
  {Damiani}, {Donati}, {Frasca}, {Jofr{\'e}}, {Lardo}, {de Laverny}, {Monaco},
  {Prisinzano}, {Sbordone}, {Sousa}, {Tautvai{\v s}ien{\.e}}, {Zaggia},
  {Zwitter}, {Delgado Mena}, {Chorniy}, {Martell}, {Silva Aguirre}, {Miglio},
  {Chiappini}, {Montalban}, {Morel}, \& {Valentini}}]{2016MNRAS.461.3336C}
{Casey}, A.~R., {Ruchti}, G., {Masseron}, T., {et~al.} 2016, \mnras, 461, 3336

\bibitem[{{Castelli} \& {Kurucz}(2003)}]{2003IAUS..210P.A20C}
{Castelli}, F., \& {Kurucz}, R.~L. 2003, in IAU Symposium, Vol. 210, Modelling
  of Stellar Atmospheres, ed. N.~{Piskunov}, W.~W. {Weiss}, \& D.~F. {Gray},
  A20

\bibitem[{{Collins} {et~al.}(2017){Collins}, {Kielkopf}, {Stassun}, \&
  {Hessman}}]{2017AJ....153...77C}
{Collins}, K.~A., {Kielkopf}, J.~F., {Stassun}, K.~G., \& {Hessman}, F.~V.
  2017, \aj, 153, 77

\bibitem[{{de Wit} {et~al.}(2013){de Wit}, {Grinin}, {Potravnov},
  {Shakhovskoi}, {M{\"u}ller}, \& {Moerchen}}]{2013A&A...553L...1D}
{de Wit}, W.~J., {Grinin}, V.~P., {Potravnov}, I.~S., {et~al.} 2013, \aap, 553,
  L1

\bibitem[{{Foster} {et~al.}(2012){Foster}, {Ji}, {Smith}, \&
  {Brickhouse}}]{2012ApJ...756..128F}
{Foster}, A.~R., {Ji}, L., {Smith}, R.~K., \& {Brickhouse}, N.~S. 2012, \apj,
  756, 128

\bibitem[{{Gingerich} {et~al.}(2017){Gingerich}, {Knapp}, {Punzi}, {Kastner},
  {Melis}, \& {Zuckerman}}]{2017AAS...22915422G}
{Gingerich}, L., {Knapp}, T., {Punzi}, K., {et~al.} 2017, in American
  Astronomical Society Meeting Abstracts, Vol. 229, American Astronomical
  Society Meeting Abstracts, 154.22

\bibitem[{{Gondoin}(2003{\natexlab{a}})}]{2003A&A...409..263G}
{Gondoin}, P. 2003{\natexlab{a}}, \aap, 409, 263

\bibitem[{{Gondoin}(2003{\natexlab{b}})}]{2003A&A...404..355G}
---. 2003{\natexlab{b}}, \aap, 404, 355

\bibitem[{{Gondoin}(2003{\natexlab{c}})}]{2003A&A...400..249G}
---. 2003{\natexlab{c}}, \aap, 400, 249

\bibitem[{{Gondoin}(2005)}]{2005A&A...444..531G}
---. 2005, \aap, 444, 531

\bibitem[{{Grinin} {et~al.}(2015){Grinin}, {Potravnov}, {Ilyin}, \&
  {Shulman}}]{2015AstL...41..407G}
{Grinin}, V.~P., {Potravnov}, I.~S., {Ilyin}, I.~V., \& {Shulman}, S.~G. 2015,
  Astronomy Letters, 41, 407

\bibitem[{{Grinin} {et~al.}(2010){Grinin}, {Potravnov}, \&
  {Musaev}}]{2010A&A...524A...8G}
{Grinin}, V.~P., {Potravnov}, I.~S., \& {Musaev}, F.~A. 2010, \aap, 524, A8

\bibitem[{{Gustafsson} {et~al.}(2008){Gustafsson}, {Edvardsson}, {Eriksson},
  {J{\o}rgensen}, {Nordlund}, \& {Plez}}]{2008A&A...486..951G}
{Gustafsson}, B., {Edvardsson}, B., {Eriksson}, K., {et~al.} 2008, \aap, 486,
  951

\bibitem[{{Herbig}(1960)}]{1960ApJ...131..632H}
{Herbig}, G.~H. 1960, \apj, 131, 632

\bibitem[{{Herbst} {et~al.}(1994){Herbst}, {Herbst}, {Grossman}, \&
  {Weinstein}}]{1994AJ....108.1906H}
{Herbst}, W., {Herbst}, D.~K., {Grossman}, E.~J., \& {Weinstein}, D. 1994, \aj,
  108, 1906

\bibitem[{{Jacoby} {et~al.}(1984){Jacoby}, {Hunter}, \&
  {Christian}}]{1984ApJS...56..257J}
{Jacoby}, G.~H., {Hunter}, D.~A., \& {Christian}, C.~A. 1984, \apjs, 56, 257

\bibitem[{{Jansen} {et~al.}(2001){Jansen}, {Lumb}, {Altieri}, {Clavel}, {Ehle},
  {Erd}, {Gabriel}, {Guainazzi}, {Gondoin}, {Much}, {Munoz}, {Santos},
  {Schartel}, {Texier}, \& {Vacanti}}]{2001A&A...365L...1J}
{Jansen}, F., {Lumb}, D., {Altieri}, B., {et~al.} 2001, \aap, 365, L1

\bibitem[{{Kaminski{\u i}} {et~al.}(2000){Kaminski{\u i}}, {Kovalchuk}, \&
  {Pugach}}]{2000ARep...44..611K}
{Kaminski{\u i}}, B.~M., {Kovalchuk}, G.~U., \& {Pugach}, A.~F. 2000, Astronomy
  Reports, 44, 611

\bibitem[{{Kastner} {et~al.}(2002){Kastner}, {Huenemoerder}, {Schulz},
  {Canizares}, \& {Weintraub}}]{2002ApJ...567..434K}
{Kastner}, J.~H., {Huenemoerder}, D.~P., {Schulz}, N.~S., {Canizares}, C.~R.,
  \& {Weintraub}, D.~A. 2002, \apj, 567, 434

\bibitem[{{Kastner} {et~al.}(2010){Kastner}, {Montez}, {Rodriguez}, {Grosso},
  {Zuckerman}, {Perrin}, {Forveille}, \& {Graham}}]{2010ApJ...719L..65K}
{Kastner}, J.~H., {Montez}, Jr., R., {Rodriguez}, D., {et~al.} 2010, \apjl,
  719, L65

\bibitem[{{Kastner} {et~al.}(2008){Kastner}, {Zuckerman}, \&
  {Forveille}}]{2008A&A...486..239K}
{Kastner}, J.~H., {Zuckerman}, B., \& {Forveille}, T. 2008, \aap, 486, 239

\bibitem[{{Kennedy} {et~al.}(2017){Kennedy}, {Kenworthy}, {Pepper},
  {Rodriguez}, {Siverd}, {Stassun}, \& {Wyatt}}]{2017RSOS....460652K}
{Kennedy}, G.~M., {Kenworthy}, M.~A., {Pepper}, J., {et~al.} 2017, Royal
  Society Open Science, 4, 160652

\bibitem[{{Mamajek}(2009)}]{2009AIPC.1158....3M}
{Mamajek}, E.~E. 2009, in American Institute of Physics Conference Series, Vol.
  1158, American Institute of Physics Conference Series, ed. T.~{Usuda},
  M.~{Tamura}, \& M.~{Ishii}, 3--10

\bibitem[{{Mamajek}(2016)}]{2016IAUS..314...21M}
{Mamajek}, E.~E. 2016, in IAU Symposium, Vol. 314, Young Stars \& Planets Near
  the Sun, ed. J.~H. {Kastner}, B.~{Stelzer}, \& S.~A. {Metchev}, 21--26

\bibitem[{{Marcy} {et~al.}(1985){Marcy}, {Duncan}, \&
  {Cohen}}]{1985ApJ...288..259M}
{Marcy}, G.~W., {Duncan}, D.~K., \& {Cohen}, R.~D. 1985, \apj, 288, 259

\bibitem[{{Melis} {et~al.}(2010){Melis}, {Gielen}, {Chen}, {Rhee}, {Song}, \&
  {Zuckerman}}]{2010ApJ...724..470M}
{Melis}, C., {Gielen}, C., {Chen}, C.~H., {et~al.} 2010, \apj, 724, 470

\bibitem[{{Melis} {et~al.}(2012){Melis}, {Zuckerman}, {Rhee}, {Song}, {Murphy},
  \& {Bessell}}]{2012Natur.487...74M}
{Melis}, C., {Zuckerman}, B., {Rhee}, J.~H., {et~al.} 2012, \nat, 487, 74

\bibitem[{{Melis} {et~al.}(2009){Melis}, {Zuckerman}, {Song}, {Rhee}, \&
  {Metchev}}]{2009ApJ...696.1964M}
{Melis}, C., {Zuckerman}, B., {Song}, I., {Rhee}, J.~H., \& {Metchev}, S. 2009,
  \apj, 696, 1964

\bibitem[{{Melis}(2009)}]{2009PhDT........34M}
{Melis}, C.~A. 2009, PhD thesis, University of California, Los Angeles

\bibitem[{{Mentuch} {et~al.}(2008){Mentuch}, {Brandeker}, {van Kerkwijk},
  {Jayawardhana}, \& {Hauschildt}}]{2008ApJ...689.1127M}
{Mentuch}, E., {Brandeker}, A., {van Kerkwijk}, M.~H., {Jayawardhana}, R., \&
  {Hauschildt}, P.~H. 2008, \apj, 689, 1127

\bibitem[{{Morrison} \& {McCammon}(1983)}]{1983ApJ...270..119M}
{Morrison}, R., \& {McCammon}, D. 1983, \apj, 270, 119

\bibitem[{{Neslu{\v s}an} \& {Budaj}(2017)}]{2017A&A...600A..86N}
{Neslu{\v s}an}, L., \& {Budaj}, J. 2017, \aap, 600, A86

\bibitem[{{Nidever} {et~al.}(2002){Nidever}, {Marcy}, {Butler}, {Fischer}, \&
  {Vogt}}]{2002ApJS..141..503N}
{Nidever}, D.~L., {Marcy}, G.~W., {Butler}, R.~P., {Fischer}, D.~A., \& {Vogt},
  S.~S. 2002, \apjs, 141, 503

\bibitem[{{Pakhomov} \& {Zhao}(2013)}]{2013AJ....146...97P}
{Pakhomov}, Y.~V., \& {Zhao}, G. 2013, \aj, 146, 97

\bibitem[{{Pecaut} \& {Mamajek}(2013)}]{2013ApJS..208....9P}
{Pecaut}, M.~J., \& {Mamajek}, E.~E. 2013, \apjs, 208, 9

\bibitem[{{Potravnov} {et~al.}(2014{\natexlab{a}}){Potravnov}, {Gorynya},
  {Grinin}, \& {Minikulov}}]{2014Ap.....57..491P}
{Potravnov}, I.~S., {Gorynya}, N.~A., {Grinin}, V.~P., \& {Minikulov}, N.~K.
  2014{\natexlab{a}}, Astrophysics, 57, 491

\bibitem[{{Potravnov} \& {Grinin}(2013)}]{2013AstL...39..776P}
{Potravnov}, I.~S., \& {Grinin}, V.~P. 2013, Astronomy Letters, 39, 776

\bibitem[{{Potravnov} {et~al.}(2014{\natexlab{b}}){Potravnov}, {Grinin},
  {Ilyin}, \& {Shakhovskoy}}]{2014A&A...563A.139P}
{Potravnov}, I.~S., {Grinin}, V.~P., {Ilyin}, I.~V., \& {Shakhovskoy}, D.~N.
  2014{\natexlab{b}}, \aap, 563, A139

\bibitem[{{Potravnov} {et~al.}(2017){Potravnov}, {Mkrtichian}, {Grinin},
  {Ilyin}, \& {Shakhovskoy}}]{2017A&A...599A..60P}
{Potravnov}, I.~S., {Mkrtichian}, D.~E., {Grinin}, V.~P., {Ilyin}, I.~V., \&
  {Shakhovskoy}, D.~N. 2017, \aap, 599, A60

\bibitem[{{Randich} {et~al.}(2001){Randich}, {Pallavicini}, {Meola},
  {Stauffer}, \& {Balachandran}}]{2001A&A...372..862R}
{Randich}, S., {Pallavicini}, R., {Meola}, G., {Stauffer}, J.~R., \&
  {Balachandran}, S.~C. 2001, \aap, 372, 862

\bibitem[{{Randich} {et~al.}(1995){Randich}, {Schmitt}, {Prosser}, \&
  {Stauffer}}]{1995A&A...300..134R}
{Randich}, S., {Schmitt}, J.~H.~M.~M., {Prosser}, C.~F., \& {Stauffer}, J.~R.
  1995, \aap, 300, 134

\bibitem[{{Rappaport} {et~al.}(2016){Rappaport}, {Gary}, {Kaye}, {Vanderburg},
  {Croll}, {Benni}, \& {Foote}}]{2016MNRAS.458.3904R}
{Rappaport}, S., {Gary}, B.~L., {Kaye}, T., {et~al.} 2016, \mnras, 458, 3904

\bibitem[{{Reddy} {et~al.}(2002){Reddy}, {Lambert}, {Laws}, {Gonzalez}, \&
  {Covey}}]{2002MNRAS.335.1005R}
{Reddy}, B.~E., {Lambert}, D.~L., {Laws}, C., {Gonzalez}, G., \& {Covey}, K.
  2002, \mnras, 335, 1005

\bibitem[{{Rieke} \& {Lebofsky}(1985)}]{1985ApJ...288..618R}
{Rieke}, G.~H., \& {Lebofsky}, M.~J. 1985, \apj, 288, 618

\bibitem[{{Ryter}(1996)}]{1996Ap&SS.236..285R}
{Ryter}, C.~E. 1996, \apss, 236, 285

\bibitem[{{Sandquist} {et~al.}(2002){Sandquist}, {Dokter}, {Lin}, \&
  {Mardling}}]{2002ApJ...572.1012S}
{Sandquist}, E.~L., {Dokter}, J.~J., {Lin}, D.~N.~C., \& {Mardling}, R.~A.
  2002, \apj, 572, 1012

\bibitem[{{Schisano} {et~al.}(2009){Schisano}, {Covino}, {Alcal{\'a}},
  {Esposito}, {Gandolfi}, \& {Guenther}}]{2009A&A...501.1013S}
{Schisano}, E., {Covino}, E., {Alcal{\'a}}, J.~M., {et~al.} 2009, \aap, 501,
  1013

\bibitem[{{Sheinis} {et~al.}(2002){Sheinis}, {Bolte}, {Epps}, {Kibrick},
  {Miller}, {Radovan}, {Bigelow}, \& {Sutin}}]{2002PASP..114..851S}
{Sheinis}, A.~I., {Bolte}, M., {Epps}, H.~W., {et~al.} 2002, \pasp, 114, 851

\bibitem[{{Shevchenko} {et~al.}(1993){Shevchenko}, {Vitrichenko}, {Grankin},
  {Ibragimov}, \& {Mel'Nikov}}]{1993AstL...19..125S}
{Shevchenko}, V.~S., {Vitrichenko}, E.~A., {Grankin}, K.~N., {Ibragimov},
  M.~A., \& {Mel'Nikov}, S.~Y. 1993, Astronomy Letters, 19, 125

\bibitem[{{Skinner} \& {G{\"u}del}(2013)}]{2013ApJ...765....3S}
{Skinner}, S.~L., \& {G{\"u}del}, M. 2013, \apj, 765, 3

\bibitem[{{Sneden}(1973)}]{1973ApJ...184..839S}
{Sneden}, C. 1973, \apj, 184, 839

\bibitem[{{Soderblom} {et~al.}(1993){Soderblom}, {Jones}, {Balachandran},
  {Stauffer}, {Duncan}, {Fedele}, \& {Hudon}}]{1993AJ....106.1059S}
{Soderblom}, D.~R., {Jones}, B.~F., {Balachandran}, S., {et~al.} 1993, \aj,
  106, 1059

\bibitem[{{Spina} {et~al.}(2014){Spina}, {Randich}, {Palla}, {Sacco},
  {Magrini}, {Franciosini}, {Morbidelli}, {Prisinzano}, {Alfaro}, {Biazzo},
  {Frasca}, {Gonz{\'a}lez Hern{\'a}ndez}, {Sousa}, {Adibekyan}, {Delgado-Mena},
  {Montes}, {Tabernero}, {Klutsch}, {Gilmore}, {Feltzing}, {Jeffries},
  {Micela}, {Vallenari}, {Bensby}, {Bragaglia}, {Flaccomio}, {Koposov},
  {Lanzafame}, {Pancino}, {Recio-Blanco}, {Smiljanic}, {Costado}, {Damiani},
  {Hill}, {Hourihane}, {Jofr{\'e}}, {de Laverny}, {Masseron}, \&
  {Worley}}]{2014A&A...567A..55S}
{Spina}, L., {Randich}, S., {Palla}, F., {et~al.} 2014, \aap, 567, A55

\bibitem[{{Stauffer} {et~al.}(1989){Stauffer}, {Hartmann}, \&
  {Jones}}]{1989ApJ...346..160S}
{Stauffer}, J.~R., {Hartmann}, L.~W., \& {Jones}, B.~F. 1989, \apj, 346, 160

\bibitem[{{Stauffer} {et~al.}(1997){Stauffer}, {Hartmann}, {Prosser},
  {Randich}, {Balachandran}, {Patten}, {Simon}, \&
  {Giampapa}}]{1997ApJ...479..776S}
{Stauffer}, J.~R., {Hartmann}, L.~W., {Prosser}, C.~F., {et~al.} 1997, \apj,
  479, 776

\bibitem[{{Stephenson}(1986)}]{1986ApJ...300..779S}
{Stephenson}, C.~B. 1986, \apj, 300, 779

\bibitem[{{Str{\"u}der} {et~al.}(2001){Str{\"u}der}, {Briel}, {Dennerl},
  {Hartmann}, {Kendziorra}, {Meidinger}, {Pfeffermann}, {Reppin}, {Aschenbach},
  {Bornemann}, {Br{\"a}uninger}, {Burkert}, {Elender}, {Freyberg}, {Haberl},
  {Hartner}, {Heuschmann}, {Hippmann}, {Kastelic}, {Kemmer}, {Kettenring},
  {Kink}, {Krause}, {M{\"u}ller}, {Oppitz}, {Pietsch}, {Popp}, {Predehl},
  {Read}, {Stephan}, {St{\"o}tter}, {Tr{\"u}mper}, {Holl}, {Kemmer}, {Soltau},
  {St{\"o}tter}, {Weber}, {Weichert}, {von Zanthier}, {Carathanassis}, {Lutz},
  {Richter}, {Solc}, {B{\"o}ttcher}, {Kuster}, {Staubert}, {Abbey}, {Holland},
  {Turner}, {Balasini}, {Bignami}, {La Palombara}, {Villa}, {Buttler},
  {Gianini}, {Lain{\'e}}, {Lumb}, \& {Dhez}}]{2001A&A...365L..18S}
{Str{\"u}der}, L., {Briel}, U., {Dennerl}, K., {et~al.} 2001, \aap, 365, L18

\bibitem[{{Turner} {et~al.}(2001){Turner}, {Abbey}, {Arnaud}, {Balasini},
  {Barbera}, {Belsole}, {Bennie}, {Bernard}, {Bignami}, {Boer}, {Briel},
  {Butler}, {Cara}, {Chabaud}, {Cole}, {Collura}, {Conte}, {Cros}, {Denby},
  {Dhez}, {Di Coco}, {Dowson}, {Ferrando}, {Ghizzardi}, {Gianotti}, {Goodall},
  {Gretton}, {Griffiths}, {Hainaut}, {Hochedez}, {Holland}, {Jourdain},
  {Kendziorra}, {Lagostina}, {Laine}, {La Palombara}, {Lortholary}, {Lumb},
  {Marty}, {Molendi}, {Pigot}, {Poindron}, {Pounds}, {Reeves}, {Reppin},
  {Rothenflug}, {Salvetat}, {Sauvageot}, {Schmitt}, {Sembay}, {Short},
  {Spragg}, {Stephen}, {Str{\"u}der}, {Tiengo}, {Trifoglio}, {Tr{\"u}mper},
  {Vercellone}, {Vigroux}, {Villa}, {Ward}, {Whitehead}, \&
  {Zonca}}]{2001A&A...365L..27T}
{Turner}, M.~J.~L., {Abbey}, A., {Arnaud}, M., {et~al.} 2001, \aap, 365, L27

\bibitem[{{Valenti} \& {Piskunov}(1996)}]{1996A&AS..118..595V}
{Valenti}, J.~A., \& {Piskunov}, N. 1996, \aaps, 118, 595

\bibitem[{{Vanderburg} {et~al.}(2015){Vanderburg}, {Johnson}, {Rappaport},
  {Bieryla}, {Irwin}, {Lewis}, {Kipping}, {Brown}, {Dufour}, {Ciardi}, {Angus},
  {Schaefer}, {Latham}, {Charbonneau}, {Beichman}, {Eastman}, {McCrady},
  {Wittenmyer}, \& {Wright}}]{2015Natur.526..546V}
{Vanderburg}, A., {Johnson}, J.~A., {Rappaport}, S., {et~al.} 2015, \nat, 526,
  546

\bibitem[{{Vogt}(1987)}]{1987PASP...99.1214V}
{Vogt}, S.~S. 1987, \pasp, 99, 1214

\bibitem[{{Vogt} {et~al.}(1994){Vogt}, {Allen}, {Bigelow}, {Bresee}, {Brown},
  {Cantrall}, {Conrad}, {Couture}, {Delaney}, {Epps}, {Hilyard}, {Hilyard},
  {Horn}, {Jern}, {Kanto}, {Keane}, {Kibrick}, {Lewis}, {Osborne},
  {Pardeilhan}, {Pfister}, {Ricketts}, {Robinson}, {Stover}, {Tucker}, {Ward},
  \& {Wei}}]{1994SPIE.2198..362V}
{Vogt}, S.~S., {Allen}, S.~L., {Bigelow}, B.~C., {et~al.} 1994, in \procspie,
  Vol. 2198, Instrumentation in Astronomy VIII, ed. D.~L. {Crawford} \& E.~R.
  {Craine}, 362

\bibitem[{{Vuong} {et~al.}(2003){Vuong}, {Montmerle}, {Grosso}, {Feigelson},
  {Verstraete}, \& {Ozawa}}]{2003A&A...408..581V}
{Vuong}, M.~H., {Montmerle}, T., {Grosso}, N., {et~al.} 2003, \aap, 408, 581

\bibitem[{{Xu} {et~al.}(2016){Xu}, {Jura}, {Dufour}, \&
  {Zuckerman}}]{2016ApJ...816L..22X}
{Xu}, S., {Jura}, M., {Dufour}, P., \& {Zuckerman}, B. 2016, \apjl, 816, L22

\bibitem[{{Yong} {et~al.}(2014){Yong}, {Roederer}, {Grundahl}, {Da Costa},
  {Karakas}, {Norris}, {Aoki}, {Fishlock}, {Marino}, {Milone}, \&
  {Shingles}}]{2014MNRAS.441.3396Y}
{Yong}, D., {Roederer}, I.~U., {Grundahl}, F., {et~al.} 2014, \mnras, 441, 3396

\bibitem[{{Zuckerman} {et~al.}(2008{\natexlab{a}}){Zuckerman}, {Fekel},
  {Williamson}, {Henry}, \& {Muno}}]{2008ApJ...688.1345Z}
{Zuckerman}, B., {Fekel}, F.~C., {Williamson}, M.~H., {Henry}, G.~W., \&
  {Muno}, M.~P. 2008{\natexlab{a}}, \apj, 688, 1345

\bibitem[{{Zuckerman} {et~al.}(2012){Zuckerman}, {Melis}, {Rhee}, {Schneider},
  \& {Song}}]{2012ApJ...752...58Z}
{Zuckerman}, B., {Melis}, C., {Rhee}, J.~H., {Schneider}, A., \& {Song}, I.
  2012, \apj, 752, 58

\bibitem[{{Zuckerman} \& {Song}(2004)}]{2004ARA&A..42..685Z}
{Zuckerman}, B., \& {Song}, I. 2004, \araa, 42, 685

\bibitem[{{Zuckerman} {et~al.}(2008{\natexlab{b}}){Zuckerman}, {Melis}, {Song},
  {Meier}, {Perrin}, {Macintosh}, {Marois}, {Weinberger}, {Rhee}, {Graham},
  {Kastner}, {Palmer}, {Forveille}, {Becklin}, {Wilner}, {Barman}, {Marcy}, \&
  {Bessell}}]{2008ApJ...683.1085Z}
{Zuckerman}, B., {Melis}, C., {Song}, I., {et~al.} 2008{\natexlab{b}}, \apj,
  683, 1085

\end{thebibliography}

\bibliographystyle{apj}

\newpage

\begin{deluxetable}{c c c}

\centering
\tablecolumns{3}
\tablewidth{320pt}
\tabletypesize{\footnotesize}

\tablecaption{\label{tbl:XRayObs}\sc XMM-Newton X-Ray Observations}

\tablehead{

	\multicolumn{1}{c}{Detector} &
	\multicolumn{1}{c}{Useful Exposure Time} &
	\multicolumn{1}{c}{Count Rate} \\

	\multicolumn{1}{c}{} &
	\multicolumn{1}{c}{(ks)} &
	\multicolumn{1}{c}{(cts s$^{-1}$)} \\
}

\startdata

pn & 19.71 & (8.3 $\pm$ 0.2) x $10^{-2}$ \\
MOS1 & 19.15 & (2.1 $\pm$ 0.1) x $10^{-2}$ \\
MOS2 & 19.16 & (2.0 $\pm$ 0.1) x $10^{-2}$ \\

\enddata
\end{deluxetable}

\newpage

\thispagestyle{empty}

\begin{deluxetable}{c c c c c c}

\centering
\tablecolumns{6}
\tablewidth{400pt}
\tabletypesize{\footnotesize}

\tablecaption{\label{tbl:OMObs}\sc XMM-Newton Optical Monitor Observations}

\tablehead{

	\multicolumn{1}{c}{Filter} &
	\multicolumn{1}{c}{Wavelength} &
	\multicolumn{1}{c}{Date} &
	\multicolumn{1}{c}{Start Time} &
	\multicolumn{1}{c}{Magnitude} &
	\multicolumn{1}{c}{Mean Magnitude} \\
	
	\multicolumn{1}{c}{} &
	\multicolumn{1}{c}{(nm)} &
	\multicolumn{1}{c}{} &
	\multicolumn{1}{c}{(UT)} &
	\multicolumn{1}{c}{(mag)} &
	\multicolumn{1}{c}{(mag)} \\
	
}

\startdata

V & 543 & 2015-01-02 & 14:38:22 & 11.52 (0.02) & \\
 & & 2015-01-02 & 15:03:44 & 11.52 (0.03) & \\
 & & 2015-01-02 & 15:29:06 & 11.51 (0.01) & \\
 & & 2015-01-02 & 15:54:28 & 11.54 (0.02) & \\
 & & 2015-01-02 & 16:19:49 & 11.51 (0.01) & \\
  & & & & & 11.514 (0.008) \\
\hline
U & 344 & 2015-01-02 & 17:15:10 & 12.368 (0.007) & \\
 & & 2015-01-02 & 17:40:32 & 12.38 (0.01) & \\
 & & 2015-01-02 & 18:05:54 & 12.366 (0.006) & \\
 & & 2015-01-02 & 18:31:16 & 12.38 (0.01) & \\
 & & 2015-01-02 & 18:56:38 & 12.371 (0.007) & \\
 & & & & & 12.371 (0.004) \\
\hline
B & 450 & 2015-01-02 & 19:21:57 & 12.38 (0.05) & \\
 & & 2015-01-02 & 19:47:20 & 12.0 (0.8) & \\
 & & 2015-01-02 & 20:12:42 & 12.38 (0.04) & \\
 & & 2015-01-02 & 21:03:25 & 12.39 (0.01) & \\
 & & & & & 12.4 (0.1) \\
\hline
UVM2 & 231 & 2015-01-02 & 21:28:45 & 15.20 (0.03) & \\
 & & 2015-01-02 & 22:39:06 & 15.20 (0.03) & \\
 & & 2015-01-02 & 23:19:27 & 15.16 (0.03) & \\
 & & 2015-01-02 & 23:59:48 & 15.22 (0.03) & \\
 & & 2015-01-03 & 00:40:09 & 15.20 (0.03) & \\
 & & & & & 15.20 (0.01) \\

\enddata

\end{deluxetable} 

\newpage

\clearpage

\newpage

\begin{deluxetable}{lcccccc}
\rotate
\tabletypesize{\normalsize}
\tablecolumns{7} 
\tablewidth{0pt}
\tablecaption{Spectroscopic Observations Summary \label{taboptspec}}
\tablehead{
 \colhead{Object} &
 \colhead{UT Date} & 
 \colhead{Instrument} & 
 \colhead{Setup} & 
 \colhead{Coverage (\AA )} & 
 \colhead{Resolution\tablenotemark{a}} & 
 \colhead{S/N \tablenotemark{b}}
}
\startdata
 RZ Psc & 14 Aug 2013 & Hamilton & 640 $\mu$m slit, Dewar \#4 & 3850-6920 & 62,000 & 60 \\
 RZ Psc & 15 Oct 2013 & Hamilton & 640 $\mu$m slit, Dewar \#4 & 3850-6920 & 62,000 & 50 \\
 RZ Psc & 16 Oct 2013 & Hamilton & 640 $\mu$m slit, Dewar \#4 & 3850-6920 & 62,000 & 50 \\
 RZ Psc & 21 Oct 2013 & HIRES & Red Collimator   & 4360-8770  & 40,000 & 100 \\
 RZ Psc & 13 Nov 2013 & Hamilton & 640 $\mu$m slit, Dewar \#4 & 3850-6920 & 62,000 & 30 \\
 RZ Psc & 14 Nov 2013 & Hamilton & 640 $\mu$m slit, Dewar \#4 & 3850-6920 & 62,000 & 35 \\
 RZ Psc & 16 Nov 2013\tablenotemark{c} & HIRES & Red Collimator   & 4360-8770 & 40,000 & 100,65,40\tablenotemark{c} \\
 RZ Psc & 21 Dec 2013 & Hamilton & 640 $\mu$m slit, Dewar \#4 & 3850-6920 & 62,000 & 40 \\
 RZ Psc & 22 Dec 2013 & Hamilton & 640 $\mu$m slit, Dewar \#4 & 3850-6920 & 62,000 & 50 \\
 RZ Psc & 03 Jan 2014 & Hamilton & 640 $\mu$m slit, Dewar \#4 & 3850-6920 & 62,000 & 60 \\
 RZ Psc & 21 Jan 2014 & Hamilton & 640 $\mu$m slit, Dewar \#4 & 3850-6920 & 62,000 & 50 \\
 RZ Psc & 22 Jan 2014 & Hamilton & 640 $\mu$m slit, Dewar \#4 & 3850-6920 & 62,000 & 40 \\
 RZ Psc & 24 Feb 2014 & Hamilton & 640 $\mu$m slit, Dewar \#4 & 3850-6920 & 62,000 & 65 \\
 RZ Psc & 28 Aug 2015 & Hamilton & 640 $\mu$m slit, Dewar \#4 & 3850-6920 & 62,000 & 65 \\
 RZ Psc & 10 Aug 2016 & Hamilton & 640 $\mu$m slit, Dewar \#4 & 3850-6920 & 62,000 & 100 \\
 J0109+2758 & 18 Nov 2016 & ESI & 1.0$''$ slit & 3900-10250 & 4000 & 60 \\
 RZ Psc & 28 Dec 2016 & Hamilton & 640 $\mu$m slit, Dewar \#4 & 3850-6920 & 62,000 & 35 \\
\enddata
\tablenotetext{a}{Resolution is measured from the FWHM of single arclines in our comparison spectra.}
\tablenotetext{b}{S/N measurement is made at 6600 \AA~in the spectrum.}
\tablenotetext{c}{Three separate spectra were obtained over the course of the night. The S/N values 
quoted are for each of the three spectra.}
\end{deluxetable} 

\newpage

\begin{flushleft}

\begin{deluxetable}{lcc}
\tabletypesize{\normalsize}
\tablecolumns{3}
\tablewidth{380pt}
\tablecaption{\sc Characteristics of RZ Psc From the Literature \label{tab:RZPscCharPrev}}
\tablehead{
 \colhead{Parameter} &
 \colhead{Value} &
 \colhead{Reference$^{a}$} \\
}
\startdata
galactic latitude & -35$^{\circ}$ &  \\
proper motion (mas/yr) & 25.4, -11.9 & 1 \\
proper motion error (mas/yr) & 2.1, 2.0 & 1 \\
T$_{eff}$ (K) & 5250 & 2 \\
& 5350$\pm$150 & 3 \\
log $g$ (cgs) & 3.41 $\pm$ 0.02 & 2 \\
& 4.2 $\pm$ 0.2 & 3 \\
$[m/H]^{b}$ & -0.3 $\pm$ 0.05 & 3 \\
v~sin\textit{i} (km s$^{-1}$) & 23 $\pm$ 1 & 2 \\
& 12.0$\pm$0.5 & 3 \\
RV (km s$^{-1}$) & -2 $\pm$ 1.5 & 4, 5 \\
& -1.2 $\pm$ 0.33 & 6 \\
& -11.75$\pm$1.1 & 2 \\
v$_{T}$ (km s$^{-1}$) & 2.0 $\pm$ 0.5 & 2 \\
& 1.0 & 3 \\
T$_{dust}$ (K) & 500 & 4 \\
$A_V$ & 0.30 $\pm$ 0.05 & 8 \\
Li 6708 \AA, EW (m\AA) & 202 $\pm$ 10 & 7 \\
H$\alpha$ EW (\AA) & 0.5 & 8 \\
L$_{IR}$/L$_{bol}$ & 0.08 & 4 \\
$R_{*}/R_{\odot}$ & 0.9 & 8 \\
$M_{*}/M_{\odot}$ & 1.0 & 8 \\
$L_{*}/L_{\odot}$ & 0.7 & 8 \\
$\dot{M}$ ($M_{\odot}$yr$^{-1}$) & $\leq$ $7 \times 10^{-12}$ & 8 \\
\enddata
\tablenotetext{a}{References: (1) Tycho2 catalog; (2) \citealt{2000ARep...44..611K}; (3) \citealt{2014A&A...563A.139P}; (4) \citealt{2013A&A...553L...1D}; (5) \citealt{1993AstL...19..125S}; (6) \citealt{2014Ap.....57..491P}; (7) \citealt{2010A&A...524A...8G}; (8) \citealt{2017A&A...599A..60P}}
\tablenotetext{b}{$[m/H]$ = log(N$_{m}$/N$_{H}$)$_{\rm star}$–log(N$_{m}$/N$_{H}$)$_{\rm Sun}$ for element m compared to hydrogen; that is, the enhancement or deficiency of an element compared to the Sun.}
\end{deluxetable}
\end{flushleft} 

\newpage

\begin{flushleft}

\begin{deluxetable}{lc}
\tabletypesize{\normalsize}
\tablecolumns{2}
\tablewidth{380pt}
\tablecaption{\sc Derived Characteristics of RZ Psc \label{tab:RZPscCharNew}}
\tablehead{
 \colhead{Parameter} &
 \colhead{Value} \\
}
\startdata
T$_{eff}$ (K) & 5600 $\pm$ 75 \\
log $g$ (cgs) & 4.35 $\pm$ 0.10\\
$[Fe/H]$ & -0.11 $\pm$ 0.03 \\
RV (km s$^{-1}$) & see Table~\ref{tabrvs} \\
v$_{T}$ (km s$^{-1}$) & 2.0 $\pm$ 0.2 \\
Li 6708 \AA, EW (m\AA) & 207 $\pm$ 6 \\
A(Li) & 3.05 \\
log(L$_{X}$/L$_{bol}$) & -3.7 to -3.2 \\
$N_{H}$ (cm$^{-2}$) & $\sim(0.39-1.83)$ x 10$^{21}$ \\
$A_{V}$ & $\sim~0.66$ \\ 
Spectral Type & G4 \\
$L_{*}/L_{\odot}$ & $\sim~0.6$ \\
\enddata
\end{deluxetable}
\end{flushleft}


\newpage

\begin{flushleft}

\begin{deluxetable}{lccccc}
	\tabletypesize{\scriptsize}
	\tablecaption{Abundances of the Elements in RZ\,Psc\label{table:abundances}}
	\tablewidth{0pt}
	\tablehead{\colhead{Species} & \colhead{Sun}& \colhead {RZ\,Psc} & \colhead{RZ\,Psc} & & \colhead{Number of}  \\
		\colhead{Species} & \colhead{A} & \colhead{A} & \colhead{[m/H]} & \colhead{$\sigma$} 
		& \colhead{Lines} 
	}
	\startdata
	
	Na I        & 6.14 & 6.18 &  0.04 & 0.09  & 4    \\
	Mg I        & 7.5  & 7.43 & --0.07 & 0.09  & 2    \\
	Al I        & 6.35 & 6.42 &  0.07 & 0.14  & 3    \\
	Si I        & 7.41 & 7.53 &  0.12 & 0.05  & 14   \\
	Ca I        & 6.24 & 6.33 &  0.09 & 0.07  & 14   \\
	Sc II       & 3.05 & 2.94 & --0.11 & 0.09  & 6    \\
	Ti I        & 4.85 & 4.81 &  0.14 & 0.08  & 25   \\
	Ti II       & 4.85 & 4.99 & --0.05 & 0.10  & 3    \\
	V I         & 3.83 & 3.99 &  0.16 & 0.08  & 11   \\
	Cr I        & 5.54 & 5.78 &  0.24 & 0.12  & 5    \\
	Mn I        & 5.33 & 5.46 &  0.13 & 0.08  & 3    \\
	Fe I        & 7.50 & 7.40 & --0.10 & 0.06  & 101  \\
	Fe II       & 7.50 & 7.41 & --0.09 & 0.07  & 13   \\
	Co I        & 4.89 & 4.93 &  0.04 & 0.15  & 4    \\
	Ni I        & 6.12 & 6.09 & --0.03 & 0.06  & 21   \\
	\enddata
\end{deluxetable}  

\end{flushleft} 

\newpage

\clearpage

\newpage

\clearpage

\newpage

\clearpage

\newpage

\clearpage

\newpage

\clearpage

\newpage

\begin{deluxetable}{lccc}
\tabletypesize{\normalsize}
\tablecolumns{4}
\tablewidth{0pt}
\tablecaption{Radial Velocities for RZ\,Psc Measured From Optical Spectra \label{tabrvs}}
\tablehead{
 \colhead{UT Date} &
 \colhead{Heliocentric Julian Date} &
 \colhead{Velocity measured} &
 \colhead{Comparison Star\tablenotemark{a}} \\
 \colhead{} &
 \colhead{(days)} &
 \colhead{(km\,s$^{-1}$)} &
 \colhead{}
}
\startdata
14 Aug 2013 & 2456610.76855 & $-$2.2$\pm$0.2 & HR 124 \\
15 Oct 2013 & 2456580.83654 & +0.0$\pm$0.2 & HR 124 \\
16 Oct 2013 & 2456581.74535 & $-$2.0$\pm$0.2 & HR 124 \\
21 Oct 2013 & 2456586.94023 & +0.4$\pm$0.2 & HR 8185 \\
13 Nov 2013 & 2456609.76353 & $-$9.2$\pm$0.5\tablenotemark{b} & HR 4027 \\
13 Nov 2013 & 2456609.76353 & +4.6$\pm$0.5\tablenotemark{b} & HR 4027 \\
14 Nov 2013 & 2456610.76855 & $-$1.8$\pm$0.2 & HR 4027 \\
16 Nov 2013 & 2456612.68576 & $-$1.6$\pm$0.2 & HR 8185 \\
16 Nov 2013 & 2456612.81636 & $-$1.7$\pm$0.2 & HR 8185 \\
16 Nov 2013 & 2456612.94820 & $-$2.4$\pm$0.2 & HR 8185 \\
21 Dec 2013 & 2456647.79631 & $-$2.0$\pm$0.2 & HR 4027 \\
22 Dec 2013 & 2456648.73979 & $-$1.8$\pm$0.2 & HR 4027 \\
03 Jan 2014 & 2456660.73687 & $-$1.9$\pm$0.2 & HR 4027 \\
21 Jan 2014 & 2456678.68097 & $-$1.7$\pm$0.2 & HR 4027 \\
22 Jan 2014 & 2456679.69084 & $-$2.0$\pm$0.2 & HR 4027 \\
24 Feb 2014 & 2456712.62651 & $-$1.7$\pm$0.2 & HR 4027 \\
28 Aug 2015 & 2457262.90766 & $-$2.3$\pm$0.2 & HR 124 \\
10 Aug 2016 & 2457610.90309 & $-$1.9$\pm$0.2 & HR 124 \\
28 Dec 2016 & 2457750.66765 & $-$1.9$\pm$0.2 & HR 124 \\
\enddata
\tablenotetext{a}{HR 8185 was observed with HIRES on UT 14 Nov 2008 in a similar setup. Otherwise, comparison stars were observed in the same night as RZ\,Psc.}
\tablenotetext{b}{Two peaks were found in the cross-correlation; see Section~\ref{sec:RV}.}
\end{deluxetable} 

\newpage

\begin{deluxetable}{c c c c}

\centering
\tablecolumns{4}
\tablewidth{400pt}
\tabletypesize{\footnotesize}

\tablecaption{\label{tbl:Models}\sc RZ\,Psc Model Parameters}

\tablehead{

        \multicolumn{1}{c}{} &
	\multicolumn{3}{c}{Model}  \\

	\multicolumn{1}{c}{Parameter} &
	\multicolumn{1}{c}{Free Abundance} &
	\multicolumn{1}{c}{T Tauri Star} &
	\multicolumn{1}{c}{Evolved Giant Star} \\
}
\startdata

$N_{H}$ (x $10^{21}$ cm$^{-2}$) & 1.8 $\pm$ 0.5 & 0.564 $\pm$ 0.008 & 0.390 $\pm$ 0.006 \\
$kT_{1}$ (keV) & 0.97 $\pm$ 0.04 & 0.67 $\pm$ 0.05 & 0.75 $\pm$ 0.04 \\
He & 1.0 & 1.0 & 0.3 \\
C & 66.6 $\pm$ 31.3 & 0.45 & 0.3 \\
N & 1.0 & 0.79 & 0.3 \\
O & 1.0 & 0.43 & 0.3 \\
Ne & 0.7 $\pm$ 1.4 & 0.83 & 1.0 \\
Mg & 1.0 & 0.26 & 0.3 \\
Al & 1.0 & 0.50 & 0.3 \\
Si & 1.0 & 0.31 & 0.3 \\
S & 1.0 & 0.42 & 0.3 \\
Ar & 1.0 & 0.55 & 0.3 \\
Ca & 1.0 & 0.195 & 0.3 \\
Fe & 0.9 $\pm$ 0.3 & 0.195 & 0.3 \\
Ne & 1.0 & 0.195 & 0.3 \\
normalization$_{1}$ (x $10^{-5}$) & 4.0 $\pm$ 1.0 & 7.6 $\pm$ 1.0 & 6.9 $\pm$ 0.6 \\
$kT_{2}$ (keV) & 0.27 $\pm$ 0.05 & 1.14 $\pm$ 0.08 & 1.6 $\pm$ 0.1 \\
normalization$_{2}$ (x $10^{-5}$) & 3.5 $\pm$ 1.0 & 6.8 $\pm$ 1.0 & 6.6 $\pm$ 0.8 \\

\enddata
\end{deluxetable} 

\newpage

\begin{deluxetable}{c c c c}

\centering
\tablecolumns{4}
\tablewidth{450pt}
\tabletypesize{\footnotesize}

\tablecaption{\label{tbl:XRaySpecAnalysis}\sc RZ Psc X-Ray Spectral Analysis}

\tablehead{

        \multicolumn{1}{c}{} &
	\multicolumn{3}{c}{Model}  \\

	\multicolumn{1}{c}{Parameter} &
	\multicolumn{1}{c}{Free Abundance} &
	\multicolumn{1}{c}{T Tauri Star} &
	\multicolumn{1}{c}{Evolved Giant Star} \\
}
\startdata

$N_{H}$ (x $10^{20}$ cm$^{-2}$) & 18.0 (5.0) & 5.64 (0.08) & 3.90 (0.06) \\
$kT_{1}$ (keV) & 0.27 (0.05) & 0.67 (0.05) & 0.75 (0.04) \\
$T_{1}$ (MK) & 3.2 (0.6) & 7.7 (0.5) & 8.8 (0.4) \\
$kT_{2}$ (keV) & 0.97 (0.04) & 1.14 (0.08) & 1.6 (0.1) \\
$T_{2}$ (MK) & 11.2 (0.4) & 13.3 (0.9) & 19.1 (1.7) \\
normalization$_{1}$ (x $10^{-5}$) & 4.0 (1.0) & 7.62 (0.09)  & 6.91 (0.07) \\
$EM_{1}$ (x $10^{52}$ cm$^{-3}$) & 2.722 & 5.253 & 4.765 \\
normalization$_{2}$ (x $10^{-5}$) & 3.5 (1.0)& 6.8 (1.0) & 6.63 (0.08) \\
$EM_{2}$ (x $10^{52}$ cm$^{-3}$) & 2.414 & 4.658 & 4.567 \\
$\chi_{red}^{2}$ & 1.053 & 1.120 & 1.298 \\
d.o.f. & 118 & 121 & 121 \\
observed flux (x $10^{-13}$ ergs s$^{-1}$ cm$^{-2}$) & 1.262 & 1.269 & 1.301 \\
intrinsic flux (x $10^{-13}$ ergs s$^{-1}$ cm$^{-2}$) & 4.152 & 1.616 & 1.653 \\
$L_{X}$ (x $10^{30}$ $\big(\frac{\textrm{D}}{170 \textrm{ pc}}\big)^{2}$ ergs s$^{-1}$) & 2.862 & 1.114 & 1.139 \\
log$(L_{X}/L_{bol})$ & (-3.230, -3.294) & (-3.640, -3.704) & (-3.630, -3.694) \\

\enddata
\end{deluxetable} 

\newpage

\clearpage

\newpage

\clearpage

\newpage

\clearpage

\end{document}